\documentclass[reprint,showpacs,aps,prd,amsfonts,
superscriptaddress,nofootinbib]{revtex4-1}

\usepackage{epsfig}
\usepackage{bm} 
\usepackage[usenames]{color}
\usepackage{bbding}

\newcommand{\Tr}{\mbox{Tr}}
\newcommand{\vphi}{\varphi}
\newcommand{\Lagr}{{\cal L}}
\newcommand{\cO}{{\cal O}}
\newcommand{\Omslash}{\!\not{\!\! \omega}}
\newcommand{\OmBarslash}{\!\not{\!\! \bar{\omega}}}

\newcommand{\bgs}{B\rightarrow X_s\gamma}
\newcommand{\gap}{\mbox{\hspace{0.5cm}}}
\newcommand{\pard}{\partial}
\newcommand{\BB}{\mbox{\boldmath$B$}}
\newcommand{\BW}{\mbox{\boldmath$W$}}
\newcommand{\BV}{\mbox{\boldmath$V$}}
\newcommand{\Bpard}{\mbox{\boldmath$\partial$}}
\newcommand{\dslash}{\!\not{\! \partial}}
\newcommand{\Bdslash}{\!\not{\! \Bpard}}
\newcommand{\Aslash}{\!\not{\!\! A}}
\newcommand{\Wslash}{\!\not{\!\! W}}
\newcommand{\BWslash}{\!\not{\!\! \BW}}
\newcommand{\Bslash}{\!\not{\!\! B}}
\newcommand{\BBslash}{\!\not{\!\! \BB}}
\newcommand{\Zslash}{\!\not{\!\! Z}}
\newcommand{\BVslash}{\!\not{\! \BV}}
\newcommand{\cL}{{\cal L}}
\newcommand{\Comm}[2]{\mbox{$[$} #1, #2\mbox{$]$}}
\newcommand{\diag}{\mbox{diag}}
\newcommand{\unit}[1]{\;\mathrm{#1}}

\begin{document}

\newcommand*{\zilina}{Physics Department, University of \v{Z}ilina, Univerzitn\'{a} 1, 010 26
\v{Z}ilina, Slovakia}
\newcommand*{\praha}{Institute of Experimental and Applied Physics,
Czech Technical University in Prague, \\
Horsk\'{a} 3a/22, 128 00 Prague, Czech Republic}
\newcommand*{\opava}{Institute of Physics, Silesian University in Opava,
Bezru\v{c}ovo n\'{a}m. 13, 746 01 Opava, Czech Republic}

\title{A 125 GeV scalar improves the low-energy data support for the top-BESS model}

\author{Mikul\'{a}\v{s} Gintner}
\email{gintner@fyzika.uniza.sk}
\affiliation{\zilina}
\affiliation{\praha}
\author{Josef Jur\'{a}\v{n}}
\email{josef.juran@utef.cvut.cz}
\affiliation{\praha}
\affiliation{\opava}

\date{\today}

\begin{abstract}
We investigate how adding a scalar resonance of a mass
$125\unit{GeV}$ affects the low-energy data support for the
top-BESS model as well as its low-energy free parameter limits.
The top-BESS model is an effective Lagrangian, a modification of
the well-known BESS model, with an ambition to describe
phenomenology of the lowest bound states of strongly-interacting
theories beyond the Standard model. In particular, the
$SU(2)_{L+R}$ vector resonance triplet of hypothetical bound
states is a centerpiece of BESS-like effective models. The
top-BESS model assumes that the triplet couples directly to the
third quark generation only. This assumption reflects a possible
special standing of the third quark generation, and the top quark
in particular, in physics of electroweak symmetry breaking. Our
findings suggest that the $125\unit{GeV}$ scalar extension of the
top-BESS model results in a higher statistical support for the
model. The best-fit values of the model's free parameters are
consistent with the top quark having a higher degree of
compositeness than the bottom quark.
\end{abstract}

\maketitle



\section{Introduction}
\label{sec:Intro}

Even though the ATLAS and CMS announcements of the 125 GeV boson
discovery~\cite{125GeVBosonDiscovery} have not settled the
question about the mechanism of electroweak symmetry breaking yet
they did provide major hints pointing to its solution. At the
moment, it is clear that the observed properties of the discovered
boson are compatible with the Standard model Higgs boson
hypothesis~\cite{EllisYou,NewTwo,NBlikeSMhiggs}. At the same time
they are compatible with many alternative extensions of the
Standard model (SM). From a theoretical point of view, the
extensions get some preference over the SM Higgs due to the
naturalness argument. They include supersymmetry theories as well
as theories where electroweak symmetry is broken by new strong
interactions, like in
Technicolor~\cite{TC,ETC,WalkingTC,TopcolorTC}.

Most studies aimed at the evaluation of the impact of the new
discovery on the alternatives theories assume the boson has a spin
zero~\cite{alternatives} even though integer spins two and
higher~\cite{LandauYang} are also admissible by the existing
experimental evidence. Of course, this assumption disfavors
strongly-interacting theories without light scalar fields and
calls for theories with a light composite strongly-interacting
Higgs~\cite{CompositeScalar-old} of a proper mass.

Following theoretical arguments, as well as the example of QCD, it
seems reasonable to expect that beside the composite scalar the
new strong interactions would also produce bound states of higher
spins. The vector $SU(2)$ triplet resonance is a natural candidate
to look for. From other point of view, in strongly interacting
theories new resonances are required to tame the unitarity. If, as
expected, the composite Higgs couplings differ from the SM ones
the Higgs alone will fail to unitarize the $VV$ $(V=W^\pm,Z)$
scattering amplitudes and other resonances are necessary to do the
job.

The new scalar can be accommodated by extra-dimensional
theories~\cite{ExtraDim} as well. There, additional new
resonances, if observed at the LHC, could be the lowest
Kaluza-Klein excitations. The attractiveness of this development
is strengthened by the Maldacena's conjecture~\cite{Maldacena} on
the dual-description relation between the extra-dimensional
weakly-interacting theories and the strongly-interacting models in
four dimensions.

The effective field theory can describe low-energy physics of
fundamental theories beyond the SM. In the case of
strongly-interacting theories where the standard tools of
perturbative field theory have limited applicability the effective
field theory is a viable tool for the phenomenology of bound
states. In addition, it can provide a unifying description of the
low-energy phenomenologies of various new physics candidates and a
useful bridge between theory and experiment. The BESS
model~\cite{BESS} is an example of the effective Lagrangian
describing the Higgsless ESB sector with an extra $SU(2)_{L+R}$
vector triplet.

Recently, we introduced and studied a modification of the BESS
model, the top-BESS model~\cite{tBESS}. The global $SU(2)_L\times
SU(2)_R$ symmetry of both models is broken down to $SU(2)_{L+R}$.
In these models, the vector resonance triplet is introduced as a
gauge field via the hidden local symmetry approach~\cite{HLS}.
Inspired by the speculations about a special role of the top quark
(or the third quark generation) in the mechanism of ESB we
modified the direct interactions of the vector triplet with
fermions. While in the BESS model there is a universal direct
coupling of the triplet to all fermions of a given chirality, in
our modification we admit direct couplings of the new triplet to
top and bottom quarks only.

In the strong scenario, the direct coupling between the SM
fermions and the vector resonance can depend on the degree of
compositeness of a given fermion as well as on symmetry group
representations the fermions are organized into. In principle, the
degree of compositeness of the SM fermions can vary for different
flavors and chirality. In addition, in some models the masses of
the SM fermions are related to the product of compositeness of the
left and right chirality~\cite{Sundrum,PomarolSerra,CHsketch}. If
so, the hierarchy of the SM fermion masses provides a reasonable
motivation to assume the third quark generation exclusivity
regarding the direct interaction with the vector resonance
triplet. On the other hand, universal degree of compositeness for
all quarks is also viable~\cite{CHsketch}.

In the top-BESS model, we allow for the possible chirality
dependence of the direct triplet-to-top/bottom coupling. In
addition, we admit disentanglement of the triplet-to-top-quark
right coupling from the triplet-to-bottom-quark right coupling.
This breaks the $SU(2)_R$ symmetry which is broken by the SM
interactions, anyway. For the sake of the top-bottom
disentanglement, we have introduced a free parameter $p$ that can
weaken, or even turn off, the strength of the triplet-to-b$_R$
coupling by assuming its value between zero and one. However, the
$SU(2)_L$ symmetry does not allow us to do the same splitting for
the left quark doublet.

Originally, as the BESS model, the top-BESS model has been
formulated without scalar resonances of any kind because the
primary goal was the systematic effective description and study of
a vector bound state physics. However, the discovery of the new
boson motivates us to study the consequences of the inclusion of a
125-GeV scalar resonance into the top-BESS model. The inclusion of
composite scalar resonance(s) into strongly interacting models
were already considered in many papers \cite{CompositeScalar-old},
\cite{CHsketch,CompositeScalar,HiggsLag,CompositeScalar-GGduality}.

In this paper, we investigate how the inclusion of the 125-GeV
scalar resonance influences the best fits of the top-BESS
parameters to the existing low-energy data. This analysis have
been performed as a multi-observable $\chi^2$-fit taking into
account the correlations among the observables used. The list of
fitted observables is comprised of $\epsilon_1$, $\epsilon_2$,
$\epsilon_3$, $\epsilon_b$ when $p=0$ and $\Gamma_b(Z\rightarrow
b\bar{b}+X)$ otherwise, and BR$(\bgs)$. The mass of the considered
vector triplet assumes TeV values. The top-BESS predictions of the
observables are derived up to a 1-loop level within some
approximations and after integrating out the vector resonance
triplet.

The maximum number of fitting parameters is four: the vector
resonance gauge coupling and three parameters, including $p$,
responsible for the direct couplings of the vector resonances to
the top and bottom quarks. We also perform fits when some of the
parameters assume fixed values. Motivation for fixing the values
can be theoretical. To simplify the analysis the free parameters
of the scalar resonance are set to their SM values.

As a by-product, this paper contains an improved analysis of the
low-energy limits of the top-BESS model without any scalar
resonances in its spectrum. In the original paper~\cite{tBESS},
the analogical analysis was based on a single-observable fits
only. Besides, here we extend the epsilon analysis by adding the
$\epsilon_2$ parameter not considered in~\cite{tBESS}.

This paper is organized as follows. Section~\ref{sec:tBESS} is
devoted to the formulation of the top-BESS model. In
Subsection~\ref{sec:tBESSnoscalar} we briefly recall its original
formulation as it has been done in~\cite{tBESS}. Then, in
Subsection~\ref{sec:AddingScalar} the model is extended by adding
a scalar field representing the 125-GeV scalar resonance. The
low-energy $\chi^2$ analysis is performed in
Section~\ref{sec:LEanalysis}. Particularly, in
Subsection~\ref{sec:pseudoobservables} we introduce observables
that will be fitted and show their relation to the anomalous
fermion couplings. In Subsection~\ref{sec:tBESSpredictions} we
provide the low-energy top-BESS expressions for the observables.
The obtained best-fit values of the free parameters of the
low-energy limit of the top-BESS model and their data support
along with the corresponding confidence intervals are shown and
discussed in several parts of Section~\ref{sec:Results}.
Conclusions are contained in Section~\ref{sec:Conclusions}. There
are three appendices in this paper. In
Appendix~\ref{app:ExpValues} the experimental values of the
observables we use in our analysis can be found.
Appendix~\ref{app:LEtBESS} is devoted to the low-energy top-BESS
Lagrangian. Finally, Appendix~\ref{app:chi2test} provides a very
brief summary of the relations used in the $\chi^2$ statistical
analysis.

\section{The top-BESS model}
\label{sec:tBESS}

The top-BESS model was originally formulated without
any new scalar fields~\cite{tBESS}.
In this section we will briefly
recall its formulation. Then, we will extend the model
by adding a scalar resonance with an ambition to
accommodate the discovery of the 125-GeV
boson.

\subsection{Original formulation: no scalar resonance}
\label{sec:tBESSnoscalar}

In this subsection we briefly summarize the top-BESS effective
Lagrangian as it was formulated in~\cite{tBESS}. The Lagrangian
possesses the $SU(2)_L\times SU(2)_R\times U(1)_{B-L}\times
SU(2)_{HLS}$ global symmetry of which the $SU(2)_L\times
U(1)_Y\times SU(2)_{HLS}$ subgroup is also a local symmetry. `HLS'
stands for the \textit{hidden local symmetry}~\cite{HLS} which is
an auxiliary gauge symmetry introduced to accommodate the $SU(2)$
triplet of vector resonances. Beside the triplet, the model
contains only the observed SM particles.

The top-BESS effective Lagrangian can be split in three parts
\begin{equation}
  \cL_\mathrm{tBESS} = \cL_\mathrm{GB} + \cL_\mathrm{ESB} + \cL_\mathrm{ferm},
  \label{eq:tBESSLag}
\end{equation}
where $\cL_\mathrm{GB}$ describes the gauge boson sector including
the $SU(2)_\mathrm{HLS}$ triplet, $\cL_\mathrm{ESB}$ is the scalar
sector responsible for spontaneous breaking of the electroweak and
hidden local symmetries, and $\cL_\mathrm{ferm}$ is the fermion
Lagrangian of the model.

In the gauge-boson sector, beside the SM gauge fields $W_\mu^a(x)$
and $B_\mu(x)$, there is the $SU(2)_\mathrm{HLS}$ gauge triplet
$\vec{V}_\mu=(V_\mu^1,V_\mu^2,V_\mu^3)$ introduced. Under the
$[SU(2)_L\times SU(2)_R]^\mathrm{glob}\times
SU(2)_{HLS}^\mathrm{loc}$ group it transforms as
\begin{equation}
  \BV_\mu \rightarrow h^\dagger \BV_\mu h +h^\dagger\pard_\mu h,
\end{equation}
where $h(x)\in SU(2)_\mathrm{HLS}^\mathrm{loc}$ and $\BV_\mu =
i\frac{g''}{2}V_\mu^a\tau^a$. The $2\times 2$ matrices
$\vec{\tau}=(\tau^1,\tau^2,\tau^3)$ are the $SU(2)$ generators.

The gauge boson Lagrangian $\cL_\mathrm{GB}$ is composed of the
Lagrangians for the individual gauge bosons
\begin{eqnarray}
   \cL_{W} &=& \frac{1}{2g^2}\Tr(\BW_{\mu\nu}\BW^{\mu\nu}),
   \\
   \cL_{B} &=& \frac{1}{2g^{\prime 2}}\Tr(\BB_{\mu\nu}\BB^{\mu\nu}),
   \\
   \cL_{V} &=& \frac{2}{g^{\prime\prime 2}}\Tr(\BV_{\mu\nu}\BV^{\mu\nu}),
\end{eqnarray}
with the field strength tensors
\begin{eqnarray}
  \BW_{\mu\nu} &=& \pard_\mu \BW_\nu - \pard_\nu \BW_\mu + \Comm{\BW_\mu}{\BW_\nu},
  \\
  \BB_{\mu\nu} &=& \pard_\mu \BB_\nu - \pard_\nu \BB_\mu,
  \\
  \BV_{\mu\nu} &=& \pard_\mu \BV_\nu - \pard_\nu \BV_\mu + \Comm{\BV_\mu}{\BV_\nu},
\end{eqnarray}
where $\BW_\mu = i g W_\mu^a\tau^a$, $\BB_\mu = i g' B_\mu Y$ are
$SU(2)_L$ and $U(1)_Y$ gauge fields.

The ESB sector contains six unphysical real scalar fields,
would-be Goldstone bosons of the model's spontaneous symmetry
breaking. Thus, naturally, the sector provides the energy scale
$v$ of ESB. The six real scalar fields $\vphi_L^a(x),
\vphi_R^a(x),\; a=1,2,3$, are introduced as parameters of
the~$SU(2)_L\times SU(2)_R$ group elements in the exp-form
$\xi(\vec{\vphi}_{L,R})=\exp(i\vec{\vphi}_{L,R}\vec{\tau}/v)\in
SU(2)_{L,R}$ where $\vec{\vphi}=(\vphi^1,\vphi^2,\vphi^3)$.

The scalar fields couple to the gauge bosons in the form given by
the $[SU(2)_L\times U(1)_Y\times SU(2)_\mathrm{HLS}]^\mathrm{loc}$
invariant Lagrangian
\begin{equation}\label{eq:LagrESB}
  \Lagr_\mathrm{ESB} = -v^2\left[\Tr\left(\bar{\omega}_\mu^\perp\right)^2
                +\alpha\Tr\left(\bar{\omega}_\mu^\parallel\right)^2\right],
\end{equation}
where $\alpha$ is a free parameter and
$\bar{\omega}_\mu^{\parallel,\perp}$ are, respectively,
$SU(2)_{L-R}$ and $SU(2)_{L+R}$ projections of the gauged
Maurer-Cartan 1-form,
\begin{eqnarray}
   \bar{\omega}_\mu^{\parallel} &=& \omega_\mu^{\parallel}+
   \frac{1}{2}\left(\xi_L^\dagger\BW_\mu\xi_L+\xi_R^\dagger\BB_\mu\xi_R\right)-
   \BV_\mu,
   \label{eq:gaugeMCparallel}\\
   \bar{\omega}_\mu^{\perp} &=& \omega_\mu^{\perp}+
   \frac{1}{2}\left(\xi_L^\dagger\BW_\mu\xi_L-\xi_R^\dagger\BB_\mu\xi_R\right),
   \label{eq:gaugeMCperp}
\end{eqnarray}
where $\omega_\mu^{\parallel,\perp}=
(\xi_L^\dagger\pard_\mu\xi_L\pm\xi_R^\dagger\pard_\mu\xi_R)/2$.

All the six scalar fields can be transformed away by an
appropriate gauge transformation. Namely, the scalar triplet
$\vec{\sigma}=(\vec{\vphi}_L+\vec{\vphi}_R)/2$ can be gauged away
by the $SU(2)_\mathrm{HLS}^\mathrm{loc}$ transformation
$h(x)=\xi(\vec{\sigma})$, leaving us with the pseudo-scalar
triplet $\vec{\pi}=(\vec{\vphi}_L-\vec{\vphi}_R)/2$. The gauge
transformation turns the Lagrangian (\ref{eq:LagrESB}) into the
gauged non-linear sigma model on the $SU(2)_L\times
SU(2)_R/SU(2)_{L+R}$ coset space. The triplet $\vec{\pi}$ plays a
role of the Goldstone bosons which supply masses to the
electroweak gauge bosons through the Higgs mechanism. The
$SU(2)_\mathrm{HLS}$ vector triplet enters the resulting
non-linear sigma model Lagrangian in the way introduced originally
by Weinberg \cite{WeinbergRho}.

The masses of the vector triplet depend on the three gauge
couplings $g, g', g''$, the free parameter $\alpha$, and the ESB
scale $v$. In the limit when $g$ and $g'$ are negligible compared
to $g''$ the masses of the neutral and charged resonances are
degenerate, $M_{V} = \sqrt{\alpha}g''v/2$. If higher order
corrections in $g/g''$ are admitted a tiny mass splitting occurs
such that $M_{V^0}>M_{V^\pm}$.

In the top-BESS model, the interactions of the new vector triplet
with the SM fermions have been modified. No new fermions beyond
the SM have been introduced. The modification singles out the new
physics role of the third quark generation, and of the top quark
in particular. Hence, the name the \textit{top-BESS model}, or
\textit{tBESS} in short.

The top-BESS fermion Lagrangian can be split in three parts
\begin{equation}
  \cL_\mathrm{ferm} = \cL_\mathrm{ferm}^\mathrm{SM} + \cL_{(t,b)}^\mathrm{tBESS}
                      + \cL_\mathrm{mass},
  \label{eq:tBESSLagFerm}
\end{equation}
where $\cL_\mathrm{ferm}^\mathrm{SM}$ is the SM part and
$\cL_{(t,b)}^\mathrm{tBESS}$ contains the modification concerning
the third quark generation
\begin{eqnarray}
  \cL_{(t,b)}^{\mbox{\scriptsize tBESS}} &=&
  b_L\left[ I_b^L(\psi_L)-I_c^L(\psi_L) \right]
  +b_R\left[ I_b^R(P\psi_R)-I_c^R(P\psi_R) \right]
  \nonumber\\
  && +2\lambda_L I_\lambda^L(\psi_L) +2\lambda_R I_\lambda^R(P\psi_R),
  \label{eq:LagrFermTBESS}
\end{eqnarray}
where $\psi=(t,b)$, and
\begin{eqnarray}
  I_c^L(\psi_L) &=& i\bar{\psi}_L(\Bdslash+\BWslash+\BBslash) \psi_L,
  \\
  I_c^R(\psi_R) &=& i \bar{\psi}_R(\Bdslash+\BBslash) \psi_R,
  \\
  I_b^h(\psi_h) &=& i\bar{\chi}_h\left[\Bdslash+\BVslash+ig'\Bslash (B-L)/2\right]\chi_h,
  \\
  I_\lambda^h(\psi_h) &=& i\bar{\chi}_h \OmBarslash^\perp \chi_h
  \nonumber\\
  &=& i \bar{\chi}_h\left[\;\Omslash^\perp+
      (\xi_L^\dagger\BWslash \xi_L-\xi_R^\dagger\BBslash^{R3} \xi_R)/2\right]\chi_h,\;\;\;\;\;\;
\end{eqnarray}
where $h=L,R$, $\BBslash^{R3}=ig'\Bslash\tau^3$, $\chi_h \equiv
\chi(\vec{\vphi}_h,\psi_h) =
\xi^\dagger(\vec{\vphi}_h)\cdot\psi_h$. The matrix $P=\diag(1,p)$
serves to disentangle the direct interaction of the vector triplet
with the right top quark from the interaction with the right
bottom quark. While $p=1$ leaves the interactions equal, the $p=0$
turns off the right bottom quark interaction completely and
maximally breaks the $SU(2)_R$ part of the Lagrangian symmetry
down to $U(1)_{R3}$.

The fermion masses are encoded in the Lagrangian term
$\cL_\mathrm{mass} = -\sum_i I_\mathrm{mass}(\psi^i)$ where the
sum runs over all SM fermion doublets and
\begin{equation}\label{eq:FermionMassTerm}
  I_\mathrm{mass}(\psi^i) = \bar{\psi}_L^i U M_f^i \psi_R^i + \mbox{H.c.},
\end{equation}
where $M_f^i$ is a $2\times 2$ diagonal matrix with the masses of
the upper and bottom fermion doublet components on its diagonal,
and
$U=\xi(\vec{\pi})\cdot\xi(\vec{\pi})=\exp(2i\vec{\pi}\vec{\tau}/v)$.

In the unitary (physical) gauge where all six unphysical scalar
fields are gauged away the gauged MC 1-form projections
(\ref{eq:gaugeMCparallel}) and (\ref{eq:gaugeMCperp}) read
\begin{eqnarray}
  \bar{\omega}_{\perp,\mu} = \frac{1}{2}(\BW_\mu-\BB_\mu), \\
  \bar{\omega}_{\parallel,\mu} = \frac{1}{2}(\BW_\mu+\BB_\mu)-\BV_\mu,
\end{eqnarray}
and the new physics part of the $(t,b)$ Lagrangian assumes the
form
\begin{eqnarray}
  \cL_{(t,b)}^\mathrm{tBESS} &=&
                            i b_L \bar{\psi}_L(\BVslash-\BWslash) \psi_L
  \nonumber\\
                 && + i b_R \bar{\psi}_R P(\BVslash-\BBslash^{R3}) P\psi_R
  \nonumber\\
  & &  + i \lambda_L \bar{\psi}_L(\BWslash-\BBslash^{R3}) \psi_L
  \nonumber\\
  & &  + i \lambda_R \bar{\psi}_RP(\BWslash-\BBslash^{R3}) P\psi_R.
  \label{eq:LagrFermTBESSinUgauge}
\end{eqnarray}

Once the gauge boson fields are expressed in the mass eigenstate
basis the mixing generated interactions of the vector triplet with
all fermions will appear on the scene. However, these indirect
interactions with the light fermions will be suppressed by the
mixing matrix elements proportional to $1/g''$.

\subsection{Adding a scalar resonance}
\label{sec:AddingScalar}

The recent discovery of the 125~GeV
boson~\cite{125GeVBosonDiscovery} motivates considerations about
possible extensions of the top-BESS model. The focus of the
pre-discovery formulation of the top-BESS model~\cite{tBESS} lain
in the study of the new physics vector resonance. To avoid
unnecessary complications the model's Lagrangian did not contain
any other non-SM fields. Now, however, the model has to face the
facts about the new 125-GeV boson.

We have chosen to investigate the simplest possibility: the
extension of the top-BESS model by the neutral scalar isoscalar
resonance field $h(x)$ of the mass $M_h=125$~GeV. Since the
transformation properties of such a field under the model's
symmetry group are trivial it is not difficult to build additional
Lagrangian terms to the top-BESS effective Lagrangian (see, e.g.,
\cite{EllisYou,CHsketch,HiggsLag}).

Adding the scalar to the top-BESS model can result in the
following modifications of the original
Lagrangian~(\ref{eq:tBESSLag})
\begin{eqnarray}
 \cL_\mathrm{ESB} &\rightarrow& \cL_\mathrm{ESB}\times
 (1+2a\frac{h}{v}+a'\frac{h^2}{v^2}+\ldots),
 \label{modif1}\\
 I_\mathrm{mass}(\psi^i)  &\rightarrow& I_\mathrm{mass}(\psi^i)
 \times(1+c_i\frac{h}{v}+c_i'\frac{h^2}{v^2}+\ldots),\;\;\;\;
 \label{modif2}
\end{eqnarray}
where $a, a', \ldots$ and $c_i, c_i', \ldots$ are free parameters.
Note that the modifications maintain the custodial symmetry. The
$a$ and $c$ parameters parameterize deviations of the $h$
couplings to the massive electroweak gauge bosons and to fermions,
respectively, from those of the SM Higgs boson. Indeed, when
$a=a'=c_i=1, \forall i$, and the rest are zeros the scalar
resonance imitates the SM Higgs boson. Of course, the mass and
kinetic terms for the scalar resonance are needed. The
self-interactions of the scalar particle can be introduced as
well.

The part of the effective Lagrangian relevant to LHC phenomenology
includes linear couplings of the scalar to SM particles and to the
vector resonance~\cite{CHsketch}. In addition, in the top-BESS
model, we will assume the flavor universality, $c_i\equiv c,
\forall i$. Then, in the unitary gauge, the top-BESS Lagrangian
with the 125 GeV scalar resonance reads
\begin{eqnarray}
 \tilde{\cL}_\mathrm{tBESS} &=& \cL_\mathrm{tBESS}
 + \frac{1}{2}\pard_\mu h\pard^\mu h - \frac{1}{2}M_h^2 h^2
 \nonumber\\
 && +\ a\left(\frac{2M_W^2}{v}W^+_\mu W^{-\mu}+\frac{M_Z^2}{v}Z_\mu Z^\mu \right.
 \nonumber\\
 && \left.
    \quad\quad\ + \frac{2M_{V^{\pm}}^2}{v}V^+_\mu V^{-\mu}+\frac{M_{V^0}^2}{v}V^0_\mu V^{0\mu}\right)h
 \nonumber\\
 && -\sum_f c\frac{m_f}{v}\left(\bar{f}_L f_R +\mbox{H.c.}\right)h,
 \label{eq:ExtendLag}
\end{eqnarray}
where $\cL_\mathrm{tBESS}$ is the original top-BESS Lagrangian and
$m_f$ stands for the fermion masses. The interactions of the
scalar resonance with all gauge fields are parametrized by the
same parameter $a$. The couplings of the scalar resonance to the
SM fermions are parameterized by $c$.

The constraints on the deviations $a$ and $c$~\cite{EllisYou}
resulting from the global analysis of the available CMS, ATLAS,
CDF, and D0 data are consistent with the 125-GeV SM Higgs boson.
Therefore, in this paper we opt for $a=c=1$ and focus on the
simplified task of fitting the vector resonance related free
parameters only.

The presence of the scalar resonance in the top-BESS particle
spectrum also affects the unitarity bounds. For the higgsless case
we have calculated such theoretical unitarity constraints in
\cite{tBESS}. As one might have expected the inclusion of the
scalar resonance can significantly relax the unitarity constraints
\cite{CHsketch}. Of course, the more closer the parameter $a$ is
to one the more relaxed the unitarity constraints are.

\section{The low-energy analysis}
\label{sec:LEanalysis}

In this section we are going to perform a multi-observable
fit of (pseudo)observables\footnote{
The quantities $\Gamma_b$ and BR$(\bgs)$ are more intimately
related to actual observables than the epsilons. To stress this
fact one might wish to nickname the epsilons as \textit{pseudo-observables}.
Nevertheless, in the following text we will not make this distinction
and will rather call the epsilons as observables, too.}
$\epsilon_1$, $\epsilon_2$, $\epsilon_3$, $\epsilon_b$ or
$\Gamma_b(Z\rightarrow b\bar{b})$,
and BR$(\bgs)$  by the top-BESS parameters.
We will study the influence
of the 125~GeV scalar resonance on the best-fit values of
the parameters and on their statistical support.
As far as the observables are concerned
the inclusion of the scalar resonance will affect
loop-level contributions to the epsilons.

\subsection{(Pseudo)observables}
\label{sec:pseudoobservables}

The epsilons are related to the \textit{basic
observables}~\cite{EpsilonMethod}: the ratio of the electroweak
gauge boson masses, $r_M\equiv M_W/M_Z$; the inclusive partial
decay width of $Z$ to the charged leptons,
$\Gamma_\ell(Z\rightarrow\ell\bar{\ell}+\mathrm{photons})$; the
forward-backward asymmetry of charged leptons at the $Z$-pole,
$A_\ell^{FB}(M_Z)$; and the inclusive partial decay width of $Z$
to bottom quarks, $\Gamma_b(Z\rightarrow b\bar{b}+X)$.

The introduction of the scalar resonance to the top-BESS model
Lagrangian does not affect the vector resonance equation of
motion. Thus, the low-energy Lagrangian obtained after integrating
out the vector resonance in~(\ref{eq:ExtendLag}) differs from the
higgsless top-BESS low-energy Lagrangian by the terms responsible
for the interactions of the scalar resonance with the SM fields
only.

The deviations of $r_M$, $\Gamma_\ell$, and $A_\ell^{FB}$ from
their predicted SM tree level values including the QED and QCD
loop contributions are parameterized by the \textit{dynamical
corrections} $\Delta r_W$, $\Delta\rho$, and $\Delta k$ as
follows~\cite{EpsilonMethod}
\begin{equation}\label{eq:rM}
  \left(1-r_M^2\right)r_M^2 =
                    \frac{\pi\alpha(M_Z)}{\sqrt{2}G_F M_Z^2 (1-\Delta r_W)}
\end{equation}
and
\begin{eqnarray}
  \Gamma_\ell &=& \frac{G_F M_Z^3}{6\pi\sqrt{2}}
                        (g_A^\ell)^2(1+r_g^2)\left(1+\frac{3\alpha}{4\pi}\right),
  \label{eq:GammaLepton}\\
     A_\ell^{FB} &=& \frac{3r_g^2}{(1+r_g^2)^2},
  \label{eq:AFBlepton}
\end{eqnarray}
where
\begin{equation}
  g_A^\ell = -\frac{1}{2}\left(1+\frac{\Delta\rho}{2}\right),
  \gap
  r_g = \frac{g_V^\ell}{g_A^\ell} = 1-4(1+\Delta k)s_0^2.
\end{equation}

The first three epsilons are defined as the combinations of the
dynamical corrections~\cite{EpsilonMethod}
\begin{eqnarray}
 \epsilon_1 &=& \Delta\rho,
 \label{eq:eps1toDynamCorrs}\\
 \epsilon_2 &=& c_0^2\Delta\rho + \frac{s_0^2}{c_{20}}\Delta r_W - 2s_0^2\Delta k,
 \label{eq:eps2toDynamCorrs}\\
 \epsilon_3 &=& c_0^2\Delta\rho + c_{20}\Delta k,
 \label{eq:eps3toDynamCorrs}
\end{eqnarray}
where $s_0$ ($c_0$) is the sine (cosine) of the Weinberg angle,
$c_{20}\equiv c_0^2-s_0^2$. The value of $s_0$ depends on the
values of $e(M_Z)$, $G_F$, and $M_Z$ through the defining relation
\begin{equation}\label{eq:s0fromGF}
 s_0^2 c_0^2 \equiv \frac{\pi\;\alpha(M_Z)}{\sqrt{2}\; G_F M_Z^2}.
\end{equation}
Given the experimental values of $\alpha(M_Z)$, $M_Z$, and $G_F$,
$s_0 $ can be considered as a replacement of $G_F$ in this list.
With this substitution the relation~(\ref{eq:rM}) can be rewritten
in a simpler form
\begin{equation}\label{eq:rMviaS0C0}
   \left(1-r_M^2\right)r_M^2 = \frac{s_0^2c_0^2}{1-\Delta r_W}.
\end{equation}

Both, the dynamical corrections and the epsilon parameters were
defined to assume zero values when the SM tree-level contributions
along with the QED and QCD loop contributions are considered only.
The SM weak loop corrections unzero the epsilon values.

The $Zbb$ vertex is naturally tested in the $Z\rightarrow
b\bar{b}+X$ decay. The corresponding decay width formula
reads~\cite{EpsilonMethod}
\begin{eqnarray}\label{eq:GammaBottom}
  \Gamma_b &=& \frac{G_F M_Z^3}{6 \pi \sqrt{2}} \beta
               \left[\frac{3-\beta^2}{2}(g_{V}^b)^2 + \beta^2 (g_{A}^b)^2\right] \times
  \nonumber\\
           &&  N_C R_{\mathrm{QCD}} \left(1+ \frac{\alpha}{12\pi} \right),
\end{eqnarray}
where $\beta = (1-4m_b^2/M_Z^2)^{1/2}$, and $R_{\mathrm{QCD}} =
1+1.2 a - 1.1a^2 -13a^3$ is the QCD correction factor, $a =
\alpha_s(M_Z)/\pi$.

The precise measurement of $\Gamma_b$ can uncover whether the
bottom quark anomalous couplings $g_{V,A}^b$ differ from the
anomalous couplings of other charged SM fermions. Assuming the
couplings differ in their $SU(2)_L$ parts only the standard
parameterization of the difference is by introducing the parameter
$\epsilon_b$~\cite{EpsilonMethod}
\begin{eqnarray}
  g_A^b &=& g_A^\ell (1+\epsilon_b),
  \label{eq:gABfromDynamicalCorrs}\\
  g_V^b &=& \left(1+\frac{\Delta\rho}{2}\right)
                    \left[-\frac{1}{2}(1+\epsilon_b)+\frac{2}{3}(1+\Delta k)s_0^2\right].
  \label{eq:gVBfromDynamicalCorrs}
\end{eqnarray}
Thus, $\epsilon_b$ along with $\epsilon_1$, $\epsilon_2$, and
$\epsilon_3$ can be used to store the experimental information
obtained by measuring the basic observables. Then, the
experimental values of the epsilons can face predictions by a
specific theory in order to transfer the information to the
parameters of the theory.

However, the top-BESS effective Lagrangian admits a more general
pattern of the bottom versus light quark anomalous coupling
difference than it is assumed in the definition of $\epsilon_b$.
Within the top-BESS model the definition's assumptions are met
when either $p=0$, or $b_R=-2\lambda_R$. Otherwise, the
experimental value of $\Gamma_b$ rather then
$\epsilon_b^{\mathrm{exp}}$ must be related to the top-BESS
theoretical prediction in order to derive the low-energy limits on
the top-BESS free parameters.

The deviations from the SM can be related to the anomalous fermion
weak couplings $\kappa$ of the fermion interaction Lagrangian
\begin{eqnarray}\label{eq:AnomFermLagr}
 \cL_\mathrm{ferm.int.}^{\mathrm{anom}} &=& -e\bar{\psi}\Aslash Q\psi
 -\frac{e}{\sqrt{2}s_0}\bar{\psi}(\Wslash^+\tau^++\Wslash^-\tau^-)\times
 \nonumber\\
 && \phantom{-e\bar{\psi}\Aslash Q\psi-}
    [(1+\kappa_L^{W\!f_{u}\!f_{d}})P_L+\kappa_R^{W\!f_{u}\!f_{d}}P_R]\psi
 \nonumber\\
 && -\frac{e}{2s_0c_0}\bar{\psi}\Zslash (K_LP_L+K_RP_R)\psi,
\end{eqnarray}
where
\begin{eqnarray}
 K_L &=& 2(T_L^3-s_0^2 Q)+
                 \diag(\kappa_L^{Z\!f_{u}\!f_{u}},\kappa_L^{Z\!f_{d}\!f_{d}}),
 \\
 K_R &=& \phantom{(T_L^3}-2s_0^2 Q\phantom{)}+
                 \diag(\kappa_R^{Z\!f_{u}\!f_{u}},\kappa_R^{Z\!f_{d}\!f_{d}}),
\end{eqnarray}
and the superscript $f_{u}$ ($f_{d}$) indicates the upper (lower)
component of any of the fermion doublets. The expressions
(\ref{eq:GammaLepton}), (\ref{eq:AFBlepton}), and
(\ref{eq:GammaBottom}) result from
\begin{eqnarray}
 \kappa_L^{Z\ell\ell} &=& -\frac{\Delta\rho}{2}
                          +2s_0^2\left[\frac{\Delta\rho}{2}
                          +\Delta k\left(1+\frac{\Delta\rho}{2}\right)\right],
 \label{eq:kappaLZllDynCorr}\\
 \kappa_R^{Z\ell\ell} &=& \phantom{-\frac{\Delta\rho}{2}+}\;\,
                           2s_0^2\left[\frac{\Delta\rho}{2}
                          +\Delta k\left(1+\frac{\Delta\rho}{2}\right)\right],
 \label{eq:kappaRZllDynCorr}\\
 \kappa_L^{Zbb} &=& g_V^b+g_A^b +1 -\frac{2}{3}s_0^2,
 \label{eq:kappaLZbbDynCorr}\\
 \kappa_R^{Zbb} &=& g_V^b-g_A^b -\frac{2}{3}s_0^2.
 \label{eq:kappaRZbbDynCorr}
\end{eqnarray}
In the case when the $\epsilon_b$ assumptions hold
\begin{eqnarray}
 \kappa_L^{Zbb} &=& -\frac{\Delta\rho}{2}(1+\epsilon_b)-\epsilon_b
 \nonumber\\
                &&  +\frac{2}{3}s_0^2\left[\frac{\Delta\rho}{2}
                    +\Delta k\left(1+\frac{\Delta\rho}{2}\right)\right],
 \label{eq:kappaLZbbDynCorrEpsb}\\
 \kappa_R^{Zbb} &=& \frac{1}{3}\kappa_R^{Z\ell\ell}.
 \label{eq:kappaRZbbDynCorrEpsb}
\end{eqnarray}

The $B\rightarrow X_s\gamma$ decay puts limits on the anomalous
$W^\pm t_{L}b_{L}$ and $W^\pm t_{R}b_{R}$
vertices~\cite{Malkawi,LariosKappaAnalysis}. In the SM it proceeds
through the flavor changing neutral current loop process
$b\rightarrow s\gamma$ dominated by the top quark exchange
diagram. The $B\rightarrow X_s\gamma$ branching fraction can be
sensitive to physics beyond the SM via new particles entering the
penguin loop. When expressed in terms of the real anomalous $Wtb$
couplings, $\kappa_L^{Wtb}$ and $\kappa_R^{Wtb}$, it can be
approximated by the following formula~\cite{LariosKappaAnalysis}
\begin{eqnarray}
 \mbox{BR}(\bgs) \times 10^{4} &=&
 3.07 + 280\,\kappa_R^{Wtb} + 2\,\kappa_L^{Wtb}
 \nonumber\\
 & &+  5520\,(\kappa_R^{Wtb})^2 + 0.3\,(\kappa_L^{Wtb})^2
 \nonumber\\
 & &+ 79\,\kappa_L^{Wtb}\kappa_R^{Wtb}.
 \label{eq:BRb2gs}
\end{eqnarray}

The experimental values of the observables used to derive the
low-energy limits on the top-BESS free parameters can be found in
Appendix~\ref{app:ExpValues}.

\subsection{Top-BESS predictions}
\label{sec:tBESSpredictions}

To derive the low-energy limits and the best-fit values of the
top-BESS free parameters we have to obtain the top-BESS
expressions for the observables. The expressions will be used to
fit the experimental values of the observables. Since the values
originate from measurements performed far below the considered
vector resonance mass, $M_V=\cO(\mathrm{TeV})$, we can work within
the framework of the low-energy limit of the top-BESS Lagrangian
obtained by integrating out the $SU(2)_{\mathrm{HLS}}$ vector
triplet.

The predictions of the low-energy top-BESS (LE-tBESS) Lagrangian
will be expressed in terms of the LE-tBESS input parameters
$\{e,s_\theta,x,M_Z,\Delta L,\Delta R,p,\{m_f\}\}$ introduced in
Appendix~\ref{app:LEtBESS} along with the LE-tBESS Lagrangian
itself.

The tree-level LE-tBESS contribution to the anomalous couplings of
the light fermions $f$ read (see Appendix~\ref{app:LEtBESS})
\begin{equation}
 \kappa_L^{W\!f_{u}\!f_{d}} = h(x;s_0)-1,
 \gap
 \kappa_R^{W\!f_{u}\!f_{d}} = 0,
\end{equation}
\begin{equation}\label{eq:kappaLRZffLE}
 \kappa_L^{Zff} = \kappa_R^{Zff} = -2s_0^2\;\Delta k^{\mathrm{LE}}(x;s_0) \;Q_f,
\end{equation}
where
\begin{equation}
 h(x;s_0) = \frac{s_0}{s_\theta}\sqrt{\frac{1+4s_\theta^2 x^2}{1+x^2}}
\end{equation}
and
\begin{equation}
 \Delta k^{\mathrm{LE}}(x;s_0) = \left(\frac{s_\theta}{s_0}\right)^2
 \frac{1+2x^2}{1+4s_\theta^2 x^2}-1.
\end{equation}
Recall that $x=g/g''$ is a free parameter of the LE-tBESS
Lagrangian (see the Eqs.~(\ref{eq:z}) and (\ref{eq:x})). Also note
that
\begin{eqnarray}
 h(x;s_0) &=& 1-s_0^2\;\Delta k^{\mathrm{LE}}(x;s_0) + \cO(x^4)
 \\
 &=& 1-0.430 \,x^2 - 0.405 \,x^4 +\ldots .
\end{eqnarray}

In the case of the top and bottom quarks the charged current
tree-level LE-tBESS anomalous couplings read
\begin{eqnarray}
 \kappa_L^{Wtb} &=& h(x;s_0)\left( 1-\frac{\Delta L}{2} \right)-1,
 \label{eq:kappaLWtbLE}\\
 \kappa_R^{Wtb} &=& h(x;s_0)\;\frac{p\;\Delta R}{2},
 \label{eq:kappaRWtbLE}
\end{eqnarray}
where the free parameters $\Delta L = b_L-2\lambda_L$ and $\Delta
R = b_R+2\lambda_R$ parameterize the top and bottom quark
couplings of the LE-tBESS Lagrangian (see the
Eqs.~(\ref{eq:CLtopbottomquarks}) -- (\ref{eq:DLRtopbottomquarks})
and (\ref{eq:DeltaLR})). The neutral current tree-level LE-tBESS
anomalous couplings read ($f=t,b$)
\begin{eqnarray}
 \kappa_L^{Zff} &=& -\Delta L\; T_L^3(f) - 2s_0^2\;\Delta k^{\mathrm{LE}}(x;s_0) \;Q_f,
 \label{eq:kappaLZQQLE}\\
 \kappa_R^{Zff} &=& \Delta R\; P_f \;T_R^3(f) - 2s_0^2\;\Delta k^{\mathrm{LE}}(x;s_0) \;Q_f,
 \label{eq:kappaRZQQLE}
\end{eqnarray}
where $P_t = 1$, $P_b = p^2$. Breaking down the compact
expressions~(\ref{eq:kappaLZQQLE}) and (\ref{eq:kappaRZQQLE})
results in
\begin{eqnarray}
 \kappa_L^{Ztt} &=& -\frac{1}{2}\Delta L -\frac{4}{3}s_0^2\;\Delta k^{\mathrm{LE}}(x;s_0) ,
 \nonumber \\
 \kappa_R^{Ztt} &=& +\frac{1}{2}\Delta R -\frac{4}{3}s_0^2\;\Delta k^{\mathrm{LE}}(x;s_0) ,
 \nonumber \\
 \kappa_L^{Zbb} &=& +\frac{1}{2}\Delta L +\frac{2}{3}s_0^2\;\Delta k^{\mathrm{LE}}(x;s_0) ,
 \nonumber \\
 \kappa_R^{Zbb} &=& -\frac{p^2}{2}\Delta R +\frac{2}{3}s_0^2\;\Delta k^{\mathrm{LE}}(x;s_0) .
 \nonumber
\end{eqnarray}

By comparing the generic anomalous
couplings~(\ref{eq:kappaLZllDynCorr}) through
(\ref{eq:kappaRZbbDynCorr}) with the relevant tree-level LE-tBESS
anomalous couplings for $f=\ell$ in~(\ref{eq:kappaLRZffLE}), and
for $f=b$ in~(\ref{eq:kappaLZQQLE}), (\ref{eq:kappaRZQQLE}), the
tree-level contributions to the dynamical corrections $\Delta\rho$
and $\Delta k$ as well as to the bottom quark vector and
axial-vector couplings $g_{V,A}^b$ can be obtained
\begin{eqnarray}
 (\Delta\rho)^{\mathrm{LE(0)}} &=& 0,
 \label{eq:DeltaRHOfromLEtBESS}\\
 (\Delta k)^{\mathrm{LE(0)}} &=& \Delta k^{\mathrm{LE}},
 \label{eq:DeltaKfromLEtBESS}
\end{eqnarray}
\begin{eqnarray}
 (g_V^b)^{\mathrm{LE(0)}} &=&
 \frac{1}{4}(\Delta L - p^2\;\Delta R)
 +\frac{2}{3}\Delta k^{\mathrm{LE}}s_0^2,
 \label{eq:gVBfromLEtBESS}\\
 (g_A^b)^{\mathrm{LE(0)}} &=&
 \frac{1}{4}(\Delta L + p^2\;\Delta R).
 \label{eq:gABfromLEtBESS}
\end{eqnarray}
The superscript $\mathrm{LE(0)}$ denotes contributions of the
LE-tBESS at tree level. Of course, by definition, only the
deviations from the corresponding SM terms can contribute
non-trivially to the dynamical corrections and to the epsilons.

In addition, the tree-level contribution to $\Delta r_W$ is
obtained when the ratio $r_M^2=(M_W/M_Z)^2$ expressed in terms of
the LE-tBESS parameters as (see Appendix~\ref{app:LEtBESS})
\begin{equation}
  r_M^2 = \frac{c_\theta^2+x^2}{(1+4s_\theta^2 x^2)(1+x^2)}
\end{equation}
is substituted to the Eq.~(\ref{eq:rMviaS0C0}). Then
\begin{equation}\label{eq:DeltaRWfromLEtBESS}
 (\Delta r_W)^{\mathrm{LE(0)}} = 1-\left(\frac{1+x^2}{1+2x^2}\right)^2.
\end{equation}

Recall that the dynamical corrections $\Delta\rho$ and $\Delta k$
can be expressed in terms of the epsilon parameters when inverting
the Eqs.~(\ref{eq:eps1toDynamCorrs}) through
(\ref{eq:eps3toDynamCorrs}). The tree-level contributions
$\epsilon_i^{\mathrm{LE}(0)}$ for $i=1,2,3$ can be obtained from
the tree-level contributions to the dynamical
corrections~(\ref{eq:DeltaRHOfromLEtBESS}),
(\ref{eq:DeltaKfromLEtBESS}), and (\ref{eq:DeltaRWfromLEtBESS}).
If $p=0$ or $\Delta R=0$, then $\epsilon_b^{\mathrm{LE}(0)}$ can
be obtained by comparing the expression~(\ref{eq:gABfromLEtBESS})
with the Eq.~(\ref{eq:gABfromDynamicalCorrs})
\begin{equation}
 \epsilon_b^{\mathrm{LE(0)}} = -\frac{1}{2}\Delta L.
\end{equation}
Otherwise, the expressions~(\ref{eq:gVBfromLEtBESS}) and
(\ref{eq:gABfromLEtBESS}) are used to calculate the LE-tBESS
tree-level value of $\Gamma_b$, employing~(\ref{eq:GammaBottom}).

Using the numerical values shown in Appendix~\ref{app:ExpValues}
the tree-level contributions to $\epsilon_i,\;i=1,2,3,$ can be
expressed as functions of $x$
\begin{eqnarray}
 {\epsilon_{1}}^{\mathrm{LE(0)}} &=& 0,
 \nonumber\\
 {\epsilon_{2}}^{\mathrm{LE(0)}} &=&
 \frac{s_0^2}{c_{20}}\frac{x^2(2+3x^2)}{(1+2x^2)^2}-2s_0^2\,\Delta k^\mathrm{LE},
 \nonumber\\
 {\epsilon_{3}}^{\mathrm{LE(0)}} &=& c_{20}\,\Delta k^\mathrm{LE}.
 \nonumber
\end{eqnarray}
Their power series read
\begin{eqnarray}
 {\epsilon_{2}}^{\mathrm{LE(0)}} &=& -2.71 \,x^4 + 2.96 \,x^6 + \ldots,
 \label{eq:eps2LE0powerseries}\\
 {\epsilon_{3}}^{\mathrm{LE(0)}} &=& x^2 + 0.66 \,x^4 + 2.56 \,x^6 + \ldots.
 \label{eq:eps3LE0powerseries}
\end{eqnarray}

There is no reason to expect that the LE-tBESS anomalies at the
tree level overwhelm the 1-loop contributions of the LE-tBESS
model to the epsilons. Thus, both contributions should be
considered when predicting the epsilon observables
\begin{equation}
 \epsilon_i^{\mathrm{LE}} = \epsilon_i^{\mathrm{LE(0)}}
                          + \epsilon_i^{\mathrm{LE(1)}},
 \gap i=1,2,3,b,
\end{equation}
where $\mathrm{LE(1)}$ denotes the 1-loop contributions of the
LE-tBESS model.

Since we study an effective non-renormalizable Lagrangian it is
not that obvious how to properly deal with the higher order
calculations~\cite{Malkawi}. One does not know the underlying
theory therefore there is no way to establish a correct scheme for
the effective Lagrangian~\cite{Georgi}. While the divergent piece
in loop calculations can be associated with a physical cutoff
$\Lambda$ up to which the effective Lagrangian is
valid~\cite{PecceiZhang}, for the finite piece there is no
completely satisfactory approach available~\cite{BurgessLondon}.

Beside the deviations in the fermion couplings of the gauge bosons
represented by the kappas, deviations from the SM in the Higgs
sector should contribute to the epsilons as well. The Higgs sector
of the scalar-less case is represented by the gauged non-linear
sigma model that can be viewed as the $M_H\rightarrow\infty$ limit
of the SM Higgs sector. In this case, we approximate
$\epsilon_i^\mathrm{LE(1)}$ by the sum of the SM weak loop
corrections $\epsilon_i^\mathrm{SM(1)}$ and the loop contributions
$\epsilon_i^{\kappa(1)}$ due to the kappas in the anomalous
fermion Lagrangian~(\ref{eq:AnomFermLagr})
\begin{equation}\label{eq:EpsilonLoopApproxHiggsless}
  \epsilon_i^{\mathrm{LE(1)}} \approx \epsilon_i^{\mathrm{SM(1)}}
                                    + \epsilon_i^{\kappa(1)}.
\end{equation}
Of course, the SM(1) contributions depend on the mass of the
non-existing SM Higgs boson. Thus, $M_H$ is just a free tuning
parameter of the approximation. We set it to the value of the
cut-off scale $\Lambda_{\mathrm{LE}}$ of the LE-tBESS Lagrangian.
The cut-off scale $\Lambda_{\mathrm{LE}}$, in turn, can be
identified with the mass of the integrated-out vector resonance.

In the 125-GeV scalar case, the first term of the
approximation~(\ref{eq:EpsilonLoopApproxHiggsless}) should be
replaced by the loop contributions due to the anomalous couplings
of the scalar resonance to the gauge bosons and to fermions.
However, since we assume that the scalar has the Higgs-like
couplings we end up with $\epsilon_i^{\mathrm{SM(1)}}$ again. This
time though $M_H$ represents the mass of the scalar resonance and
should be set to $125\unit{GeV}$.

To cover both possibilities --- no scalar as well as the 125-GeV
scalar resonance
--- the 1-loop SM contributions have been evaluated for
four different values of $M_H$. The 125~GeV value corresponds to
the case when a new scalar resonance imitating the LHC discovery
is added to the spectrum of the top-BESS model; 300~GeV, 1~TeV,
and 2~TeV values correspond to the top-BESS model without a scalar
resonance.

The 1-loop SM contributions to the epsilon parameters are given by
the following relations~\cite{newBESS}
\begin{eqnarray}
  \epsilon_1^{\mathrm{SM(1)}} &=& \left(+5.60-0.86\ln\frac{M_H}{M_Z}\right)\times 10^{-3},
  \label{eq:eps1SM}\\
  \epsilon_2^{\mathrm{SM(1)}} &=& \left(-7.09+0.16\ln\frac{M_H}{M_Z}\right)\times 10^{-3},
  \label{eq:eps2SM}\\
  \epsilon_3^{\mathrm{SM(1)}} &=& \left(+5.25+0.54\ln\frac{M_H}{M_Z}\right)\times 10^{-3},
  \label{eq:eps3SM}\\
  \epsilon_b^{\mathrm{SM(1)}} &=&       -6.43\times 10^{-3}.
  \label{eq:epsbSM}
\end{eqnarray}
The numerical values of $\epsilon_i^{\mathrm{SM(1)}}$ for the
quoted values of $M_H$ are shown in Table~\ref{tab:epsSMvalues}.
\begin{table}
\caption{The values of the SM 1-loop contributions to the epsilon
         parameters considering four different masses of the SM
         Higgs boson: 125~GeV, 300~GeV, 1~TeV, and 2~TeV.}
 \label{tab:epsSMvalues}
\begin{ruledtabular}
  \begin{tabular}{ccccc}
     $M_H$ (GeV) & 125 & 300 & 1000 & 2000 \\
     \hline
     $\epsilon_1^{\mathrm{SM(1)}}\times 10^{3}$ & \phantom{-}5.33
     & \phantom{-}4.58 & \phantom{-}3.54 & \phantom{-}2.94 \\
     $\epsilon_2^{\mathrm{SM(1)}}\times 10^{3}$ & -7.04 & -6.90
     & -6.71 & -6.60 \\
     $\epsilon_3^{\mathrm{SM(1)}}\times 10^{3}$ & \phantom{-}5.42
     & \phantom{-}5.89 & \phantom{-}6.54 & \phantom{-}6.92 \\
     $\epsilon_b^{\mathrm{SM(1)}}\times 10^{3}$ & -6.43 & -6.43 & -6.43 & -6.43 \\
  \end{tabular}
\end{ruledtabular}
\end{table}

The 1-loop SM contributions to $g_V^b$ and $g_A^b$ can be obtained
subtracting the SM tree-level couplings from the SM tree plus
1-loop couplings
\begin{equation}
 (g_{V,A}^b)^{\mathrm{SM(1)}} =
 (g_{V,A}^b)^{\mathrm{SM(0+1)}}-
 (g_{V,A}^b)^{\mathrm{SM(0)}},
\end{equation}
where $(g_{V,A}^b)^{\mathrm{SM(0+1)}}$ are given by the
Eqs.~(\ref{eq:gABfromDynamicalCorrs}) and
(\ref{eq:gVBfromDynamicalCorrs}) if $\Delta\rho=(\Delta
\rho)^{\mathrm{SM(1)}}$, $\Delta k=(\Delta k)^{\mathrm{SM(1)}}$,
and $\epsilon_b=\epsilon_b^{\mathrm{SM(1)}}$ are applied. Of
course, $(g_{V}^b)^{\mathrm{SM(0)}}=-1/2+2s_0^2/3$ and
$(g_{A}^b)^{\mathrm{SM(0)}}=-1/2$.

The $\epsilon_i^{\kappa(1)}$ contributions can be calculated using
the results of~\cite{LariosKappaAnalysis}. In the context of the
non-linear electroweak chiral Lagrangian the paper provides
expressions for new physics loop contributions to the epsilon
parameters in terms of generic anomalous couplings of the
Lagrangian~(\ref{eq:AnomFermLagr}). Up to the order of
$m_t^2\ln\Lambda^2$ the anomalous loop contributions
read~\cite{LariosKappaAnalysis}
\begin{eqnarray}
  \epsilon_1^{\mathrm{NP(1)}} &=& \frac{3 m_t^2 G_F}{2 \sqrt{2} \pi^2}\ln\frac{\Lambda^2}{m_t^2}
                       \left[ \kappa_L^{Wtb}\left( 1+{\kappa_L^{Wtb}}\right)\right.
  \nonumber\\
                   & & +
  \left. \left(\kappa_R^{Ztt}-\kappa_L^{Ztt}\right)\left(1-\kappa_R^{Ztt}+\kappa_L^{Ztt}\right) \right],
  \\
  \epsilon_2^{\mathrm{NP(1)}} &=& \epsilon_3^{\mathrm{NP(1)}} \;=\; 0\,,
  \\
  \epsilon_b^{\mathrm{NP(1)}} &=& \frac{m_t^2 G_F}{2 \sqrt{2} \pi^2} \ln\frac{\Lambda^2}{m_t^2}
  \nonumber\\
  && \times
  \left[\left(\kappa_L^{Ztt}-\frac{1}{4}\kappa_R^{Ztt}\right)\left(1+2\kappa_L^{Wtb}\right)\right],
\end{eqnarray}
where $\Lambda$ is the cut-off scale of the effective Lagrangian
under consideration. In the cases when the $\mathrm{NP(1)}$
contributions depend on $\kappa^{Zbb}$ the dependence is
suppressed by $m_b\ll m_t$.

To obtain $\epsilon_1^{\kappa(1)}$ and $\epsilon_b^{\kappa(1)}$
one has to substitute the LE-tBESS anomalous
couplings~(\ref{eq:kappaLWtbLE}) -- (\ref{eq:kappaRZQQLE}) into
the expressions above. The cut-off scale of the LE-tBESS
Lagrangian has been set to the mass of the integrated-out vector
resonance, $\Lambda = \Lambda_{\mathrm{LE}} = M_V$. When
$M_V=1$~TeV and using the numerical values of
Appendix~\ref{app:ExpValues} the leading terms of the $x^2$ series
of $\epsilon_1^{\kappa(1)}$ and $\epsilon_b^{\kappa(1)}$ read
\begin{eqnarray}
 \frac{\epsilon_1^{\kappa(1)}}{10^{-2}} &=& \phantom{-}6.57 \,\Delta R
 - 2.82 \,(2 - 3\,\Delta L) \, x^2 + \ldots,
 \label{eq:eps1dLE1series}\\
 \frac{\epsilon_b^{\kappa(1)}}{10^{-2}} &=& -0.55\, (4\, \Delta L + \Delta R)
 \nonumber\\
  & & - (1.88 - 0.47\, \Delta R  - 3.76\, \Delta L)\, x^2 + \ldots,
  \label{eq:epsbdLE1series}
\end{eqnarray}
where we have also neglected non-linear terms in $\Delta L$,
$\Delta R$. The $\epsilon_1^{\kappa(1)}$ and
$\epsilon_b^{\kappa(1)}$ series for $M_V=0.3$~TeV and $M_V=2$~TeV
are obtained when multiplying (\ref{eq:eps1dLE1series}) and
(\ref{eq:epsbdLE1series}) by the numerical factors of $0.31$ and
$1.39$, respectively.

In our analysis, we have not calculated $(g_{V,A}^b)^{\kappa(1)}$.
Thus, the fit when $p\neq 0$ is based on the $\mathrm{LE(0)}$ and
$\mathrm{SM(1)}$ contributions to $g_{V,A}^b$ only. We justified
this simplifying approximation by comparing the single-observable
fits based on
$\epsilon_b^{\mathrm{LE(0)}+\mathrm{SM(1)}+\kappa(1)}$ with the
fits based on $(g_{V,A}^b)^{\mathrm{LE(0)}+\mathrm{SM(1)}}$ when
$p=0$, see \cite{tBESS}. Fig.~11 of \cite{tBESS} illustrates that
the absence of the $\kappa(1)$ contribution in the latter fits
introduces only relatively small shifts in the obtained confidence
level contours.

The LE-tBESS prediction of $\mbox{BR}(\bgs)$ has been calculated
by substituting (\ref{eq:kappaLWtbLE}) and (\ref{eq:kappaRWtbLE})
into (\ref{eq:BRb2gs}).

\section{Results}
\label{sec:Results}

Using the exact formulas for the LE-tBESS predictions
we have performed a multi-parameter $\chi^2$ fit of the
observables in order to obtain the most preferred values
and confidence level intervals for the LE-tBESS parameters. Two
slightly different sets of observables have been fitted.
The first set of fitted observables consists of
$\{\epsilon_1,\epsilon_2,\epsilon_3,\epsilon_b,\mbox{BR}(\bgs)\}\equiv$
Set-A. The fit of the Set-A has been used to investigate the
low-energy data support for the top-BESS model when the vector
resonance does not couple to the right bottom quark, i.e. $p=0$.
The second set of fitted observables consists of
$\{\epsilon_1,\epsilon_2,\epsilon_3,\Gamma_b,\mbox{BR}(\bgs)\}\equiv$
Set-B. In the fit of the Set-B, $p$ can assume non-zero values
and it can be one of the free fitting parameters of the top-BESS
model. The experimental values of the observables are
shown in Appendix~\ref{app:ExpValues}. The basic $\chi^2$-fit
relations used in our calculations are summarized in
Appendix~\ref{app:chi2test}.

\subsection{Single- vs. multi-observable fits}
\label{sec:SingleVsMulti}

In~\cite{tBESS} we have derived the limits on the top-BESS
parameters as the intersections of various single-observable fits.
Here, we perform comparison with the multi-observable fit results
based on the same observables.

We start by restricting the LE-tBESS parameter $x$. Since $g''$ is
a function\footnote{ Assuming that the values of $G_F$, $e(M_Z)$,
and $M_Z$ are fixed by measurement, the explicit formula for
$g''(x)$ is given by the Eqs.~(\ref{eq:gx}) and (\ref{eq:gppx}). }
of $x$, this translates into limits on the most fundamental of the
top-BESS free parameters: the $SU(2)_{\mathrm{HLS}}$ gauge
coupling $g''$. The physically sensible values of $g''$ are
bounded from below by the unitarity limits and from above by the
perturbativity limit, $g''/2\lesssim 4\pi$. The unitarity limits
were studied in \cite{tBESS} and depend on the mass of the new
vector triplet resonance. If we require that the top-BESS model
unitarity holds up to the same energy as for the Higgsless SM ---
$1.7\unit{TeV}$
--- the $g''$ parameter is restricted only from below:
$g''\geq 3$, $6$, and $9$, when $M_V = 1.0$, $1.7$, and
$2.3\unit{TeV}$, respectively. The perturbativity limit reads
$g''\lesssim 30$.

The Eqs.~(\ref{eq:eps2LE0powerseries}) and
(\ref{eq:eps3LE0powerseries}), along with the expressions for
$\epsilon_i^{\mathrm{SM}(1)}$ and $\epsilon_i^{\kappa(1)}$ imply
that $\epsilon_3$ is the most sensitive epsilon with respect to
$x$. Hence, our first estimation of the limit on $x$ comes from
$\epsilon_3$ alone.

The LE-tBESS prediction of $\epsilon_3$ can be approximated by
\begin{equation}\label{eq:eps3LEtBESSprediction}
 \epsilon_3=x^2+\epsilon_3^{\mathrm{SM(1)}}(M_H),
\end{equation}
where $M_H$ is either identified with the cut-off scale
$\Lambda_{\mathrm{LE}}=M_V$, if the original top-BESS model
without a scalar resonance is considered, or it is a mass of the
scalar resonance added to the top-BESS model. We have
compared~(\ref{eq:eps3LEtBESSprediction}) with
$\epsilon_3^{\mathrm{exp}}$ for four different values of $M_H$:
$125$~GeV, $300$~GeV, $1$~TeV, and $2$~TeV. While the first value
suits the top-BESS model with the scalar resonance imitating the
recently discovered 125-GeV boson, the other three values can
represent the top-BESS model without a scalar resonance.

\begin{figure}
\includegraphics[scale=0.75]{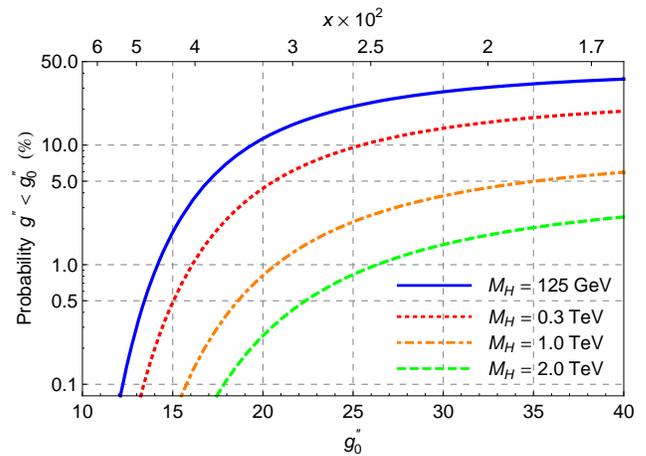}
\caption{\label{fig:gppLimitFromEps3}(color online) The
probability that $g''$ lies anywhere below a given value of
$g_0''$. It is based on $\epsilon_3$ parameter and depends on
$M_H$ used for calculation of $\epsilon_3^{\mathrm{SM(1)}}$. Plots
for $M_H=125$~GeV (blue solid), $0.3$~TeV (red dotted), $1$~TeV
(orange dot-dashed), and $2$~TeV (green dashed) are shown. Recall
that $x=g/g''$.}
\end{figure}

In all these cases, $\epsilon_3^{\mathrm{SM(1)}}(M_H)$ is larger
than $\epsilon_3^{\mathrm{exp}}$. Thus, the
Eq.~(\ref{eq:eps3LEtBESSprediction}) has no real solution for $x$.
Nevertheless, the positive values of the difference are
statistically admissible if we assume its normal distribution with
the standard deviation taken from $\epsilon_3^\mathrm{exp}$. Then,
the probability that the difference is positive amounts to $47\%$,
$28\%$, $10\%$, and $5\%$ when $M_H=0.125$, $0.3$, $1$, and
$2$~TeV, respectively. At the same time, these numbers indicate
the data support for $g''$ taking on any real value. The
likelihood that the $g''$ value lies anywhere below a given value
$g_0''$ is depicted in Fig.~\ref{fig:gppLimitFromEps3}. We can see
that adding the 125-GeV scalar field to the top-BESS Lagrangian
improves the data support for the model.

Generally speaking, this simplified single-observable analysis
provides not very convincing statistical support of the BESS-like
extensions of the SM. Nevertheless, two comments seem to be in
order. First, the negativity of
$\epsilon_3^\mathrm{exp}-\epsilon_3^{\mathrm{SM(1)}}$ could be
compensated for by adding new independent direct interactions of
the vector triplet with the light fermions. It would work in the
same way as the $b$ parameter does in the BESS model~\cite{BESS}.
Secondly, as it will be shown below, the muti-observable fitting
modifies the $\epsilon_3$-based conclusions.

\begin{table}
\caption{The best-fitting values of the free parameters
$\{g''(x)$, $\Delta L$, $\Delta R\}$ (Set-A) and $\{g''(x)$,
$\Delta L$, $\Delta R$, $p\}$ (Set-B) obtained by fitting five
observables at various fixed values of $\Lambda$ that correspond
to the Higgsless top-BESS model ($M_H=\Lambda$). The
$\chi^2_\mathrm{min}$ and the corresponding statistical backing of
the optimized values are also shown.}
 \label{tab:FitHiggsless}
\begin{ruledtabular}
\begin{tabular}{ccccccc}
  \multicolumn{7}{c}{Set-A ($\mathrm{d.o.f.}=2$)} \\
  $\Lambda$ (TeV) & $g''(x)$ & $\Delta L$ & $\Delta R$  &    & $\chi^2_{min}$ & Backing (\%) \\
  \hline
  0.3 & 77       & -0.003 & 0.092 & &   6.14 & 4.6 \\
  1   & $\infty$ & -0.003 & 0.052 & &   7.06 & 2.9 \\
  2   & $\infty$ & -0.003 & 0.048 & &   8.03 & 1.8 \\
  \hline
  \multicolumn{7}{c}{Set-B ($\mathrm{d.o.f.}=1$)} \\
  $\Lambda$ (TeV) & $g''(x)$ & $\Delta L$ & $\Delta R$ & $p$ & $\chi^2_{min}$ & Backing (\%) \\
  \hline
  0.3 & 75       & -0.004 & 0.092 & 0.036 & 2.78 & 9.5 \\
  1   & $\infty$ & -0.006 & 0.052 & 0.063 & 3.70 & 5.4 \\
  2   & $\infty$ & -0.007 & 0.048 & 0.069 & 4.67 & 3.1 \\
\end{tabular}
\end{ruledtabular}
\end{table}

\begin{table}
\caption{The same as in Table~\ref{tab:FitHiggsless} except that
the 125-GeV scalar is added to the top-BESS model ($M_H =
125\unit{GeV}$).}
 \label{tab:Fit125GeV}
\begin{ruledtabular}
\begin{tabular}{ccccccc}
  \multicolumn{7}{c}{Set-A ($\mathrm{d.o.f.}=2$)} \\
  $\Lambda$ (TeV) & $g''(x)$ & $\Delta L$ & $\Delta R$ & & $\chi^2_{min}$ & Backing (\%) \\
  \hline
  1 & 29 & -0.003 & 0.016 &   & 5.78 & 5.6 \\
  2 & 29 & -0.003 & 0.011 &   & 5.78 & 5.6 \\
  \hline
  \multicolumn{7}{c}{Set-B ($\mathrm{d.o.f.}=1$)} \\
  $\Lambda$ (TeV) & $g''(x)$ & $\Delta L$ & $\Delta R$ & $p$ & $\chi^2_{min}$ & Backing (\%) \\
  \hline
  1 & 29 & -0.004 & 0.016 & 0.209 & 2.40 & 12.1 \\
  2 & 29 & -0.004 & 0.011 & 0.289 & 2.40 & 12.1 \\
\end{tabular}
\end{ruledtabular}
\end{table}

Let us turn our attention to the multi-observable fitting. More
observables not only bring additional experimental input but they
also make the analysis sensitive to the other-than-$x$ free
parameters. The results of the multi-observable fits are shown in
Tables~\ref{tab:FitHiggsless} and \ref{tab:Fit125GeV}. When
fitting the Set-A, we search for the values of three LE-tBESS free
parameters, $x$, $\Delta L$, $\Delta R$, that would best fit the
observables $\epsilon_{1,2,3,b}$, and $\mbox{BR}(\bgs)$. When
fitting the Set-B we calculate the best fits of four LE-tBESS
parameters, $x$, $\Delta L$, $\Delta R$, $p$, to the observables
$\epsilon_{1,2,3}$, $\Gamma_b$, and $\mbox{BR}(\bgs)$. Recall
that, because of $\epsilon_b$, fitting of the Set-A is applicable
only when $p=0$. There is a one-to-one relationship between $x$
and $g''$; in Tables~\ref{tab:FitHiggsless} and
\ref{tab:Fit125GeV} we list the optimized values of $g''$, rather
than $x$.

Table~\ref{tab:FitHiggsless} corresponds to the case of the
Higgsless top-BESS model. There, $M_H$ can be considered an
adjusting parameter of the used approximation. Its value is set to
$\Lambda$. The best-fit values of $g''$ obtained for this case are
very high with quite a low statistical support. In this sense, the
results get worse as the value of $\Lambda=M_V$ is raised. For the
Set-B, the preferred value of $p$ is pushed close to zero. Thus,
the results of the Set-A fit and the Set-B fit are almost
identical.

Table~\ref{tab:Fit125GeV} shows the optimized values of the
LE-tBESS parameters with the 125-GeV resonance added. When
compared with the values of Table~\ref{tab:FitHiggsless} we
observe a significant decrease of the best value of $g''$. At the
same time, its backing has improved. The multi-observable analysis
confirms the trend deduced from the fit of $\epsilon_3$ alone: the
presence of the 125-GeV scalar boson in the top-BESS model moves
$g''$ to lower values. The optimized value of $p$ ranges within
$0.2$ -- $0.3$. Thus, data prefers the right bottom-quark coupling
to the vector resonance triplet to be less than $10\%$ of the
corresponding right top-quark coupling. This finding is in a
general agreement with the expectations of some partial
compositeness hypotheses that the degree of compositeness of the
top quark is higher than it is for the bottom
quark~\cite{CHsketch}.

\begin{figure}
\includegraphics[scale=0.70]{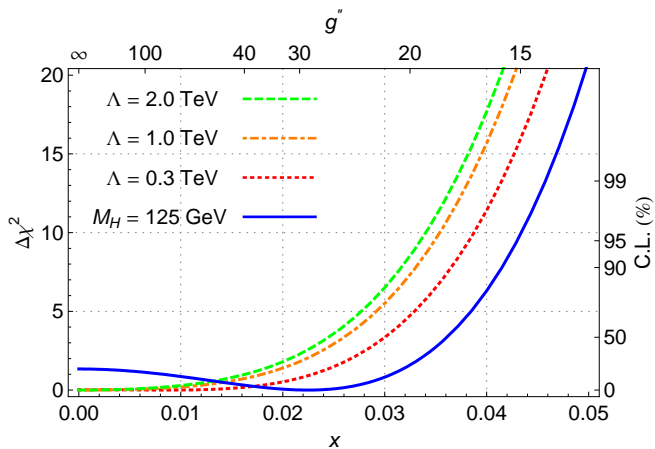}
\caption{\label{fig:chi2Multiobs}(color online)
 $\Delta\chi^2$ as a function of $x$ when
 $\Delta L$, $\Delta R$, and $p$ are
 fixed at their optimized values found in the Set-B fit
 (see Tables ~\ref{tab:FitHiggsless} and \ref{tab:Fit125GeV}).
 The blue solid curve corresponds to the top-BESS model
 with the 125-GeV scalar boson and $\Lambda = 1\unit{TeV}$.
 The red dotted, orange dot-dashed, and green dashed curves
 correspond to the Higgsless top-BESS model with
 $\Lambda=0.3$, $1$, and $2\unit{TeV}$, respectively.
 The right-hand side $y$-axis depicts the values of
 the four-dimensional confidence level limits that correspond
 to the $\Delta\chi^2$ labeled on the left-hand side $y$-axis.}
\end{figure}

In Fig.~\ref{fig:chi2Multiobs} we show how
$\Delta\chi^2=\chi^2(x,\Delta L,\Delta
R,p;\Lambda,M_H)-\chi^2_{min}$ depends on $x$ (and thus on $g''$)
when $\Delta L$, $\Delta R$, and $p$ are kept at their optimized
values found in the Set-B fit (see Tables~\ref{tab:FitHiggsless}
and \ref{tab:Fit125GeV}). On the right-hand side $y$-axis the
confidence level values for the four-dimensional parameter space
are labeled. Thus, for a chosen probability, we can read off the
confidence level interval for $x$ (or $g''$) in the $(\Delta
L,\Delta R,p)$-optimized slice of the four-dimensional parameter
space. There are three curves corresponding to the Higgsless case
($\Lambda = 0.3$, $1$, and $2\unit{TeV}$) and the curve of the
125-GeV scalar case. While in the former there is a clear
preference of very large (infinite) $g''$, the latter prefers
$g''$ of about 30. The 125-GeV scalar curve depicted in
Fig.~\ref{fig:chi2Multiobs} has been calculated at
$\Lambda=1\unit{TeV}$. The curve is barely distinguishable from
the $\Lambda=2\unit{TeV}$ case; hence, we do not show it.

\subsection{Fitting of sub-models}
\label{sec:FittingOfSubmodels}

In this Subsection we analyze and compare the fits of the
observables when the values of some of the free parameters are
fixed. Recall that if the Set-A is fitted by the LE-tBESS
Lagrangian there are three fitting parameters, $\{x, \Delta L,
\Delta R\}$. In the case of fitting the Set-B, there are four
fitting parameters, $\{x, \Delta L, \Delta R, p\}$. However, some
beyond the SM theories can result in the top-BESS effective
description where one or more of these parameters are given.

We saw that in the case of the Higgsless top-BESS model the
experiment pushes the value of $g''$ to infinity which would
effectively remove the vector resonance from the game. In spite of
this preference, we have performed fits when the values of $g''$
are fixed at $10$, $20$, and $30$, and the cut-off scale $\Lambda$
runs over $0.3$, $1$, and $2\unit{TeV}$. The obtained best-fit
values of the parameters $\{\Delta L,\Delta R,p\}$ and their
backings are shown in Table~\ref{tab:SetBdof2Higgsless}. As we can
see, for the fixed values of $g''$ the backing improves when the
cut-off scale decreases. Unfortunately, it also implies lowering
the value of the vector resonance mass which can conflict with the
exclusion lower bounds obtained by the LHC.

\begin{table}
\caption{The best-fit values of the $\{\Delta L, \Delta R, p\}$
 parameters in the Set-B fits of the Higgsless case ($M_H=\Lambda$)
 for various values of $g''$ and $\Lambda$.
 The corresponding values of $\chi^2_{min}$
 and of statistical backing for $\mathrm{d.o.f}=5-3=2$ are also shown.
}
 \label{tab:SetBdof2Higgsless}
\begin{ruledtabular}
\begin{tabular}{ccccccc}
  \multicolumn{7}{c}{Set-B ($\mathrm{d.o.f}=2$)} \\
  $\Lambda$ (TeV)  & $g''$ & $\Delta L$ & $\Delta R$ & $p$ & $\chi^2_{min}$ & Backing (\%) \\
  \hline
0.3   &       10 & -0.006 & 0.290 & 0.011 &           26.1 & \phantom{1}0.0 \\
0.3   &       20 & -0.005 & 0.135 & 0.024 & \phantom{1}4.1 & 13.1 \\
0.3   &       30 & -0.005 & 0.109 & 0.030 & \phantom{1}3.0 & 22.4 \\
  1   &       10 & -0.008 & 0.111 & 0.030 &           33.8 & \phantom{1}0.0 \\
  1   &       20 & -0.007 & 0.067 & 0.049 & \phantom{1}6.7 & \phantom{1}3.6 \\
  1   &       30 & -0.006 & 0.059 & 0.056 & \phantom{1}4.6 & \phantom{1}9.8 \\
  2   &       10 & -0.009 & 0.090 & 0.037 &           38.6 & \phantom{1}0.0 \\
  2   &       20 & -0.007 & 0.058 & 0.057 & \phantom{1}8.6 & \phantom{1}1.4 \\
  2   &       30 & -0.007 & 0.052 & 0.063 & \phantom{1}6.0 & \phantom{1}4.9
\end{tabular}
\end{ruledtabular}
\end{table}

\begin{figure}
\includegraphics[scale=0.85]{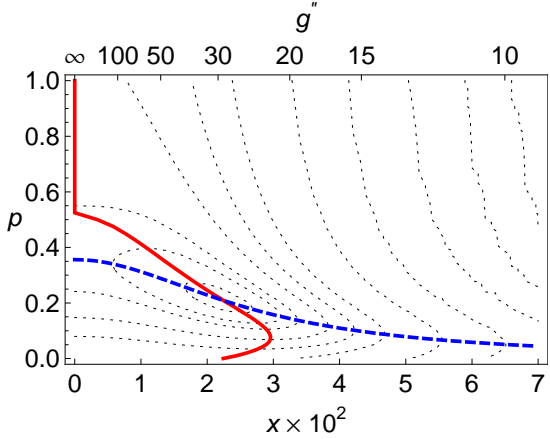}
\caption{\label{fig:pxMap}(color online)
  The graph depicts the results of various Set-B fits
  when $M_H=125$~GeV and $\Lambda=1$~TeV.
  The blue dashed line shows the best-fitting value
  of $p$ as a function of $x$ with
  $\{\Delta L, \Delta R, p\}$ as fitting parameters.
  The red solid line is the best-fitting value of $x$
  as a function of $p$
  with $\{\Delta L, \Delta R, x\}$ as fitting parameters.
  The gray dotted contours join the $(x,p)$ points
  for which the Set-B fit with
  $\{\Delta L, \Delta R\}$ as fitting parameters provides
  the same backing.}
\end{figure}

We also saw that the inclusion of the 125-GeV scalar resonance
into the top-BESS model has created much better situation:
$g''\approx 30$ rather than the infinity is preferred by the data.
Now, let us perform the fit when either $x$ or $p$ are fixed. In
the former case, the fitting parameters are $\{\Delta L, \Delta R,
p\}$. The dashed line in Fig.~\ref{fig:pxMap} shows how the
best-fit value of $p$ depends on the fixed $x$. We can see that
the preferred value of $p$ ranges between $0$ and $0.4$. Thus, the
vector resonance interaction with the right bottom quark weaker
than with the right top quark is supported by the data. In the
same graph, the solid line depicts how the best-fit value of $x$
depends on the fixed $p$ when the fitting parameters are $\{\Delta
L, \Delta R, x\}$. The preferred values of $g''$ do not fall below
$20$.

In Fig.~\ref{fig:pxMap}, there are also the gray dotted contours
which join the $(x,p)$ points with the same backing in the fit
with $\{\Delta L, \Delta R\}$ as fitting parameters. As expected,
the $(x,p)$ point with the greatest backing in this fit
corresponds to the intersection of the dashed and solid lines.

In Table~\ref{tab:SetBdof2Scalar125}, we show the best-fit values
of the parameters $\{\Delta L,\Delta R,p\}$ and the corresponding
backings for the 125-GeV scalar case. The values of $g''$ are
fixed at $10$, $20$, and $30$ and the cut-off scale $\Lambda$ is
either $1\unit{TeV}$ or $2\unit{TeV}$. The best data support at
the level of $30\%$ ($\mathrm{d.o.f.=2}$) is obtained when
$g''=30$. Note that there is a difference between the
$1\unit{TeV}$ and $2\unit{TeV}$ cases only in the best-fit values
of $\Delta R$ and $p$.

\begin{table}
\caption{The same as in Table~\ref{tab:SetBdof2Higgsless}
         except that the 125-GeV scalar case
         ($M_H=125\unit{GeV}$) is considered here.}
 \label{tab:SetBdof2Scalar125}
\begin{ruledtabular}
\begin{tabular}{ccccccc}
  \multicolumn{7}{c}{Set-B ($\mathrm{d.o.f}=2$)} \\
  $\Lambda$ (TeV)  & $g''$ & $\Delta L$ & $\Delta R$ & $p$ & $\chi^2_{min}$ & Backing (\%) \\
  \hline
  1   &       10 & -0.005 & 0.065 & 0.051 & 21.2           & \phantom{1}0.0 \\
  1   &       20 & -0.004 & 0.023 & 0.144 & \phantom{1}2.8 & 24.5 \\
  1   &       30 & -0.004 & 0.015 & 0.215 & \phantom{1}2.4 & 30.1 \\
  2   &       10 & -0.005 & 0.047 & 0.070 & 21.2           & \phantom{1}0.0 \\
  2   &       20 & -0.004 & 0.017 & 0.198 & \phantom{1}2.8 & 24.5 \\
  2   &       30 & -0.004 & 0.011 & 0.298 & \phantom{1}2.4 & 30.1
\end{tabular}
\end{ruledtabular}
\end{table}

The only observables that depend on $\Delta R$ and $p$ are
$\Gamma(\bgs)$ and $\Gamma_b$, while $\epsilon_1$ depends solely
on $\Delta R$. In particular, $\Gamma(\bgs)$ depends on the
product $p\,\Delta R$ while $\Gamma_b$ depends on $p^2\Delta R$.
The sensitivity to $\Lambda$ enters through $\epsilon_1$ only.
Scrutinizing Tables~\ref{tab:SetBdof2Higgsless} and
\ref{tab:SetBdof2Scalar125} one can find that the values of the
product $p\,\Delta R$ differ only slightly through all lines:
$3.19\leq p\,\Delta R\times 10^3\leq 3.33$
(Table~\ref{tab:SetBdof2Higgsless}) and $3.28\leq p\,\Delta
R\times 10^3\leq 3.37$ (Table~\ref{tab:SetBdof2Scalar125}). Thus,
it seems that $\epsilon_1$ and $\Gamma(\bgs)$ dominate the
determination of the best-fit values of $\Delta R$ and $p$.

\begin{figure}
\includegraphics[scale=0.75]{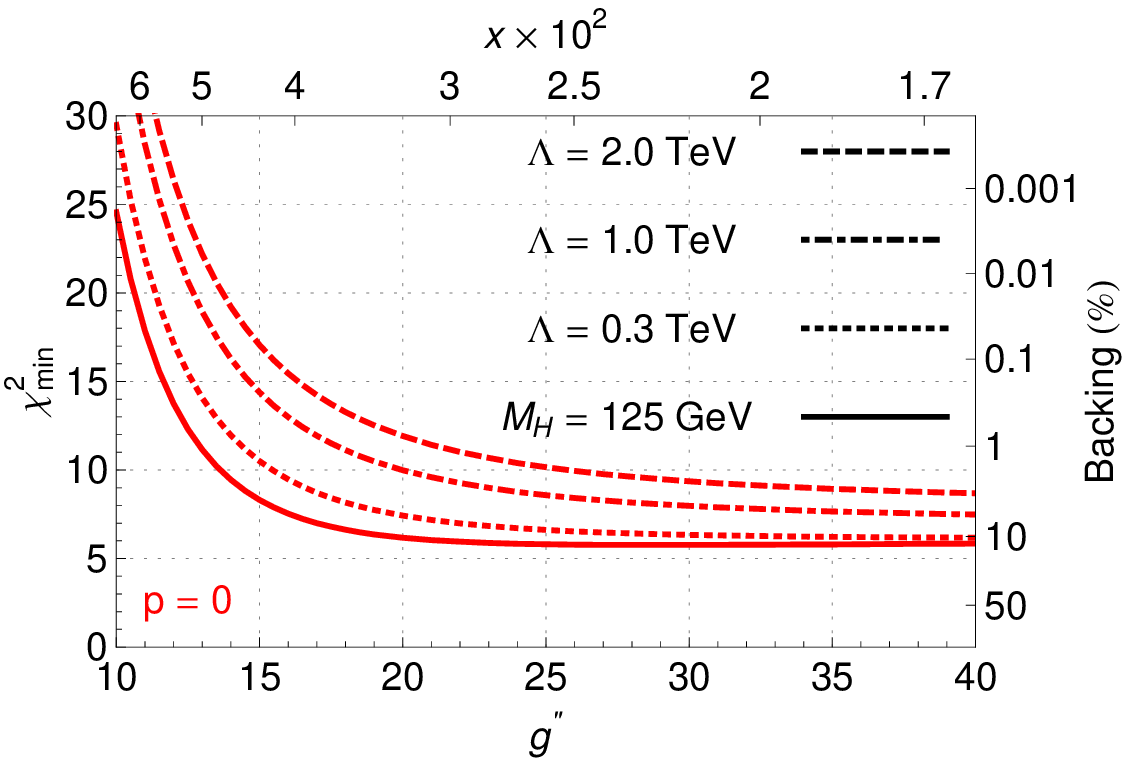}
\includegraphics[scale=0.75]{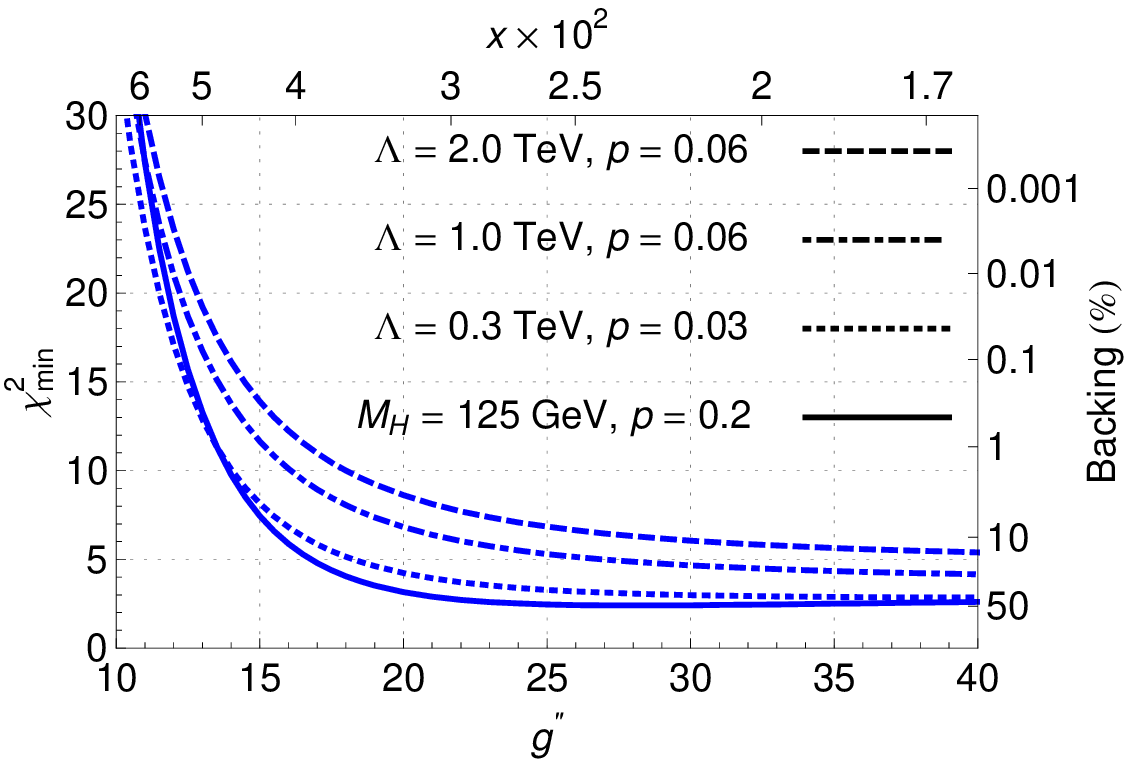}
\caption{\label{fig:chi2minDOF2gpp}(color online)
 $\chi^2_{min}$ of the Set-B fit as a function of $g''$
 with $\{\Delta L, \Delta R\}$ as fitting parameters
 $(\mathrm{d.o.f.}=3)$.
 The solid line corresponds to the 125-GeV scalar case
 with $M_H=125\unit{GeV}$ and $\Lambda=1\unit{TeV}$.
 The dotted, dot-dashed, and dashed lines correspond
 to the Higgsless cases of $M_H=\Lambda=0.3$, $1$, and
 $2\unit{TeV}$, respectively.
 The top panel corresponds to $p=0$. In the bottom pannel,
 the $p$ values are shown in the graph's legend.
 The right-hand $y$ axis labels indicate the backing for
 $\mathrm{d.o.f.}=3$.
 Some of the best-fit values of $\Delta L$ and $\Delta R$
 can be found in Tables~\ref{tab:SetABdof3Higgsless}
 and~\ref{tab:SetABdof3Scalar125}.
}
\end{figure}

\begin{figure}
\includegraphics[scale=0.75]{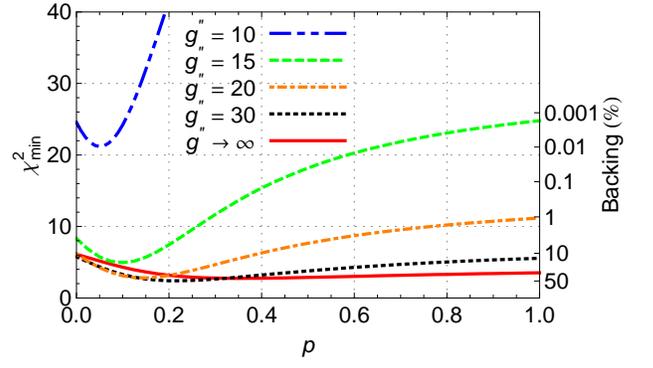}
\caption{\label{fig:chi2minDOF2p}(color online)
 $\chi^2_{min}$ of the Set-B fit as a function of $p$
 with $\{\Delta L, \Delta R\}$ as fitting parameters
 $(\mathrm{d.o.f.}=3)$.
 The 125-GeV scalar case
 with $M_H=125\unit{GeV}$ and $\Lambda=1\unit{TeV}$ is considered only.
 The blue dot-dot-dashed, green dashed, orange dot-dashed,
 black dotted, and red solid curves correspond to $g''=10, 15, 20, 30$,
 and $g''\rightarrow\infty$, respectively.
 Some of the best-fit values of $\Delta L$ and $\Delta R$
 can be found in Table~\ref{tab:SetABdof3Scalar125}.
 }
\end{figure}

Figs.~\ref{fig:chi2minDOF2gpp} and \ref{fig:chi2minDOF2p}
illustrate how $\chi^2_\mathrm{min}$ depends on $g''$ and $p$ in
the two-parameter fits of the Set-B when the fitting parameters
are $\{\Delta L, \Delta R\}$. In both Figures, the right-hand $y$
axis labels indicate the backings for $\mathrm{d.o.f.}=5-2=3$.


In Fig.~\ref{fig:chi2minDOF2gpp}, we show $\chi^2_\mathrm{min}$ as
a function of $g''$ for both top-BESS models, the Higssless one as
well as the 125-GeV scalar one. The former with various values of
the cut-off scale $\Lambda$ and the latter with
$\Lambda=1\unit{TeV}$. While the top panel contain curves when the
direct vector resonance interaction to the right bottom quark is
turned off, $p=0$, in the bottom panel the $p$ values are fixed
close to the best-fit values found in the three-parameter
$\{\Delta L,\Delta R, p\}$ fit (see
Tables~\ref{tab:SetBdof2Higgsless} and
\ref{tab:SetBdof2Scalar125}). In Fig.~\ref{fig:chi2minDOF2p},
there are the $\chi^2_\mathrm{min}$ curves as function of $p$ for
the 125-GeV scalar top-BESS model with $\Lambda=1\unit{TeV}$ only.
Again, we can see that the inclusion of the 125-GeV scalar into
the top-BESS model improves the model's agreement with the
low-energy precision data. The best-fit values of $\{\Delta L,
\Delta R\}$ and the corresponding backings when
$\mathrm{d.o.f.}=3$ for selected values of the given parameters
$g''$ and $p$ are shown in Tables~\ref{tab:SetABdof3Higgsless} and
\ref{tab:SetABdof3Scalar125}.

\begin{table}
\caption{The best-fit values of the $\{\Delta L, \Delta R\}$
         parameters in the Set-A and Set-B fits of the Higgsless
         case ($M_H=\Lambda$) for various values of $g''$, $p$, and $\Lambda$.
         The corresponding values of $\chi^2_{min}$
         and of statistical backing for $\mathrm{d.o.f}=5-2=3$ are also shown.
}
 \label{tab:SetABdof3Higgsless}
\begin{ruledtabular}
\begin{tabular}{ccccccc}
  \multicolumn{7}{c}{Set-A ($\mathrm{d.o.f}=3$)} \\
  $\Lambda$ (TeV) & $g''$ &     & $\Delta L$ & $\Delta R$ & $\chi^2_{min}$ & Backing (\%) \\
  \hline
  1   &       10 &   & -0.005 & 0.111 & 37.2           & 0.0 \\
  1   &       20 &   & -0.004 & 0.067 & 10.0           & 1.8 \\
  1   &       30 &   & -0.003 & 0.059 & \phantom{1}8.0 & 4.6 \\
  \hline
  \multicolumn{7}{c}{Set-B ($\mathrm{d.o.f}=3$)} \\
  $\Lambda$ (TeV) & $g''$ & $p$ & $\Delta L$ & $\Delta R$ & $\chi^2_{min}$ & Backing (\%) \\
  \hline
0.3     &       10 & 0.03  & -0.006 & 0.253 & 34.3           & \phantom{1}0.0 \\
0.3     &       20 & 0.03  & -0.005 & 0.131 & \phantom{1}4.2 & 23.7 \\
0.3     &       30 & 0.03  & -0.005 & 0.109 & \phantom{1}3.0 & 39.3 \\
  1     &       10 & 0     & -0.008 & 0.111 & 37.1           & \phantom{1}0.0 \\
  1     &       10 & 0.06  & -0.008 & 0.107 & 37.4           & \phantom{1}0.0 \\
  1     &       20 & 0     & -0.007 & 0.067 & 10.0           & \phantom{1}1.9 \\
  1     &       20 & 0.06  & -0.007 & 0.066 & \phantom{1}6.8 & \phantom{1}7.8 \\
  1     &       30 & 0     & -0.006 & 0.059 & \phantom{1}8.0 & \phantom{1}4.7 \\
  1     &       30 & 0.06  & -0.006 & 0.058 & \phantom{1}4.7 & 19.8 \\
  2     &       10 & 0.06  & -0.009 & 0.089 & 40.1           & \phantom{1}0.0 \\
  2     &       20 & 0.06  & -0.007 & 0.058 & \phantom{1}8.6 & \phantom{1}3.5 \\
  2     &       30 & 0.06  & -0.007 & 0.052 & \phantom{1}6.1 & 10.9
\end{tabular}
\end{ruledtabular}
\end{table}

\begin{table}
\caption{The same as in Table~\ref{tab:SetABdof3Higgsless}
         except that the 125-GeV scalar case
         ($M_H=125\unit{GeV}$) is considered here.}
 \label{tab:SetABdof3Scalar125}
\begin{ruledtabular}
\begin{tabular}{ccccccc}
  \multicolumn{7}{c}{Set-A ($\mathrm{d.o.f}=3$)} \\
  $\Lambda$ (TeV) & $g''$ &     & $\Delta L$ & $\Delta R$ & $\chi^2_{min}$ & Backing (\%) \\
  \hline
  1   &       10 &   & -0.004 & 0.065 & 24.6 & \phantom{1}0.0 \\
  1   &       20 &   & -0.003 & 0.023 & \phantom{1}6.2 & 10.3 \\
  1   &       30 &   & -0.003 & 0.015 & \phantom{1}5.8 & 12.3 \\
  \hline
  \multicolumn{7}{c}{Set-B ($\mathrm{d.o.f}=3$)} \\
  $\Lambda$ (TeV) & $g''$ & $p$ & $\Delta L$ & $\Delta R$ & $\chi^2_{min}$ & Backing (\%) \\
  \hline
  1     &       10 & 0     & -0.005 & 0.065 & 24.6           & \phantom{1}0.0 \\
  1     &       10 & 0.2   & -0.006 & 0.045 & 41.6           & \phantom{1}0.0 \\
  1     &       20 & 0     & -0.004 & 0.023 & \phantom{1}6.2 & 10.3 \\
  1     &       20 & 0.2   & -0.004 & 0.021 & \phantom{1}3.2 & 36.8 \\
  1     &       30 & 0     & -0.004 & 0.015 & \phantom{1}5.8 & 12.3 \\
  1     &       30 & 0.2   & -0.004 & 0.016 & \phantom{1}2.4 & 49.1 \\
  2     &       10 & 0.25  & -0.006 & 0.035 & 38.4           & \phantom{1}0.0 \\
  2     &       30 & 0.25  & -0.004 & 0.012 & \phantom{1}2.5 & 48.1 \\
\end{tabular}
\end{ruledtabular}
\end{table}

Using Tables~\ref{tab:SetABdof3Higgsless} and
\ref{tab:SetABdof3Scalar125} along with
Tables~\ref{tab:FitHiggsless} and \ref{tab:Fit125GeV} we can see
the level of consistency between the fits based on the Set-A and
Set-B observables. We attribute the small discrepancies to not
including the $(g_{V,A}^b)^{\kappa(1)}$ contributions when
calculating $\Gamma_b$ of the Set-B observables as it was
mentioned at the end of Subsection~\ref{sec:tBESSpredictions}.

\begin{figure}
\centering
\includegraphics[scale=0.80]{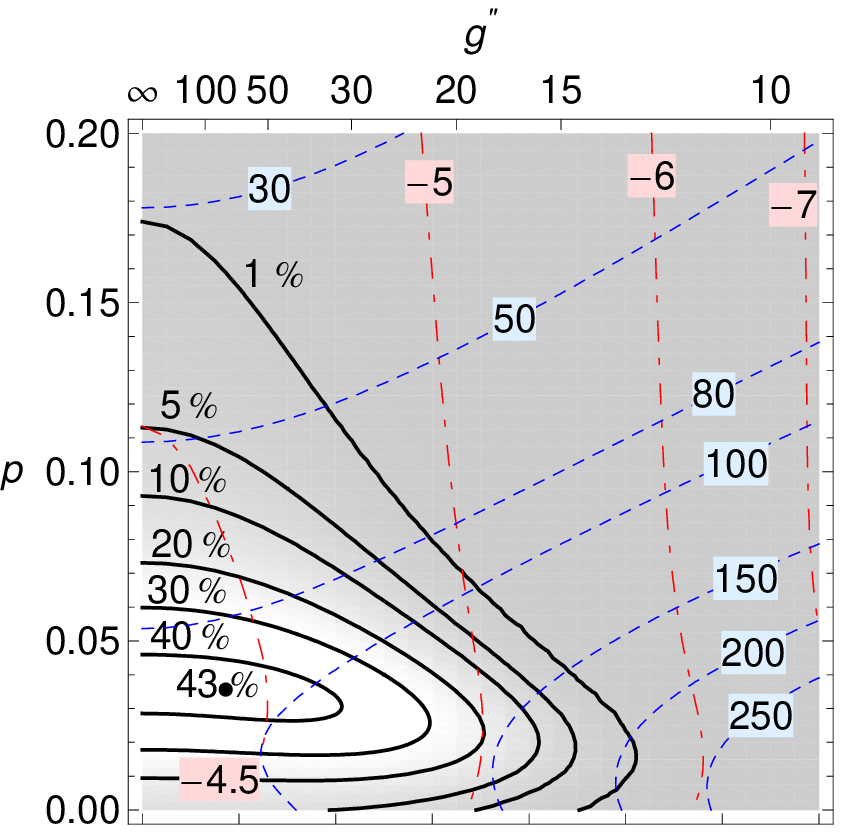}
\includegraphics[scale=0.818]{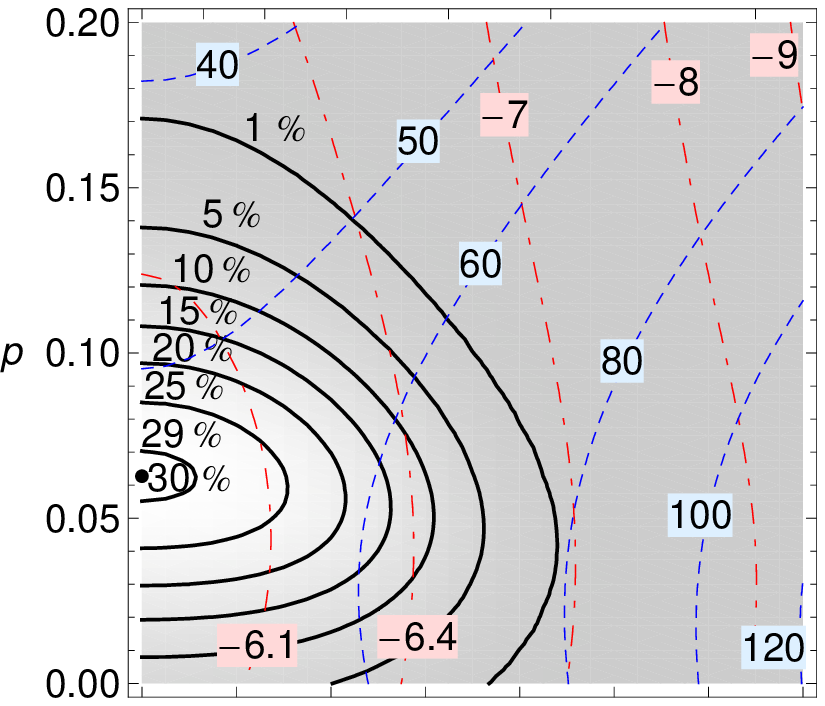}
\includegraphics[scale=0.80]{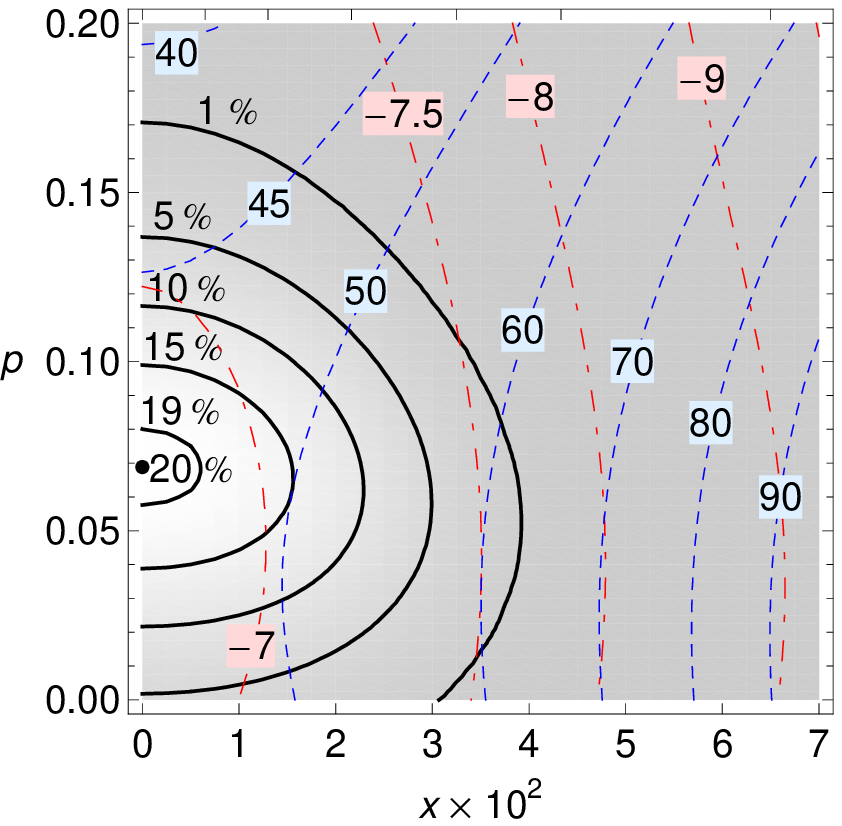}
\caption{\label{fig:ContourMapsHiggsless}(color online)
         The contour maps of backings as functions of $x$
         and $p$ for the Set-B fits of the Higgsless
         case ($M_H=\Lambda$) with $\{\Delta L, \Delta R\}$
         as fitting parameters ($\mathrm{d.o.f.} = 3$).
         The three panels correspond to $\Lambda=0.3$,
         $1$, and $2\unit{TeV}$, from the top to the bottom,
         respectively. The black solid contours connect
         $(x,p)$ points of the same backing. The backing
         is also indicated by the shading of
         the background: the lighter gray means greater
         backing. The best-fit values of $\Delta L$ and
         $\Delta R$ for the given $(x,p)$ point can be
         read off from the dashed-line grid. The constant
         values of $\Delta L$ and $\Delta R$ are represented
         by the red dot-dashed and blue dashed lines,
         respectively. The values are equal to the numbers
         attached to the grid lines divided by $10^3$.
 }
\end{figure}

\begin{figure*}
\centering
\includegraphics[scale=0.75]{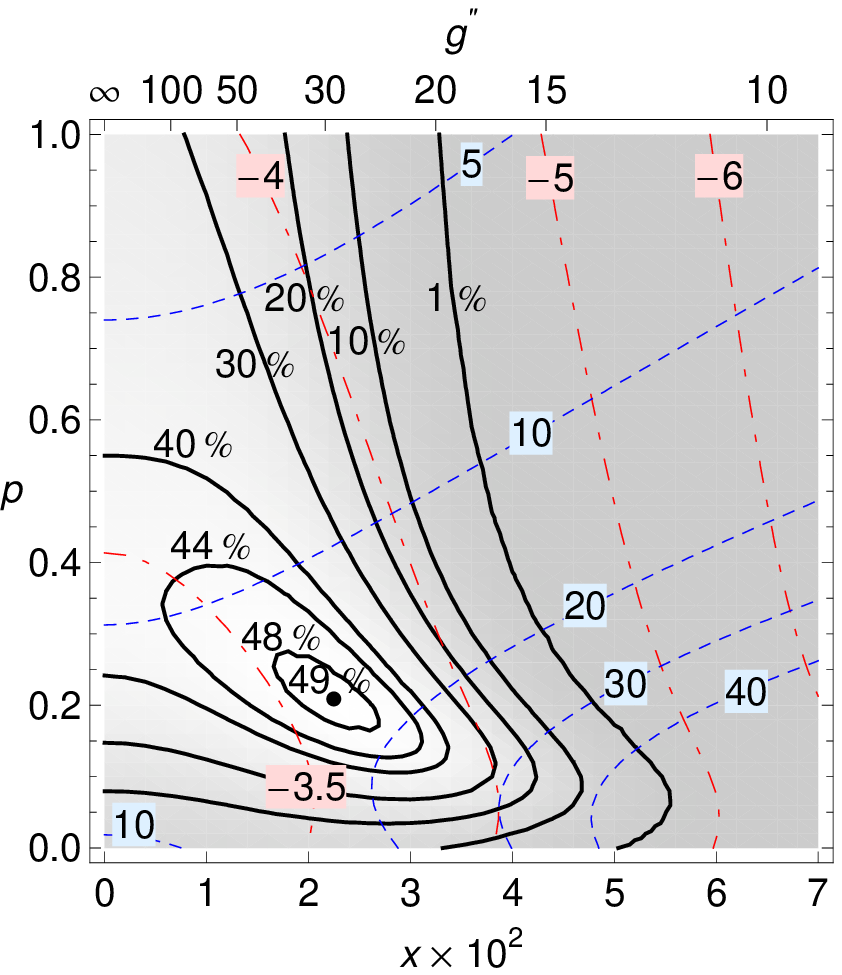}
\hspace{0.5cm}
\includegraphics[scale=0.75]{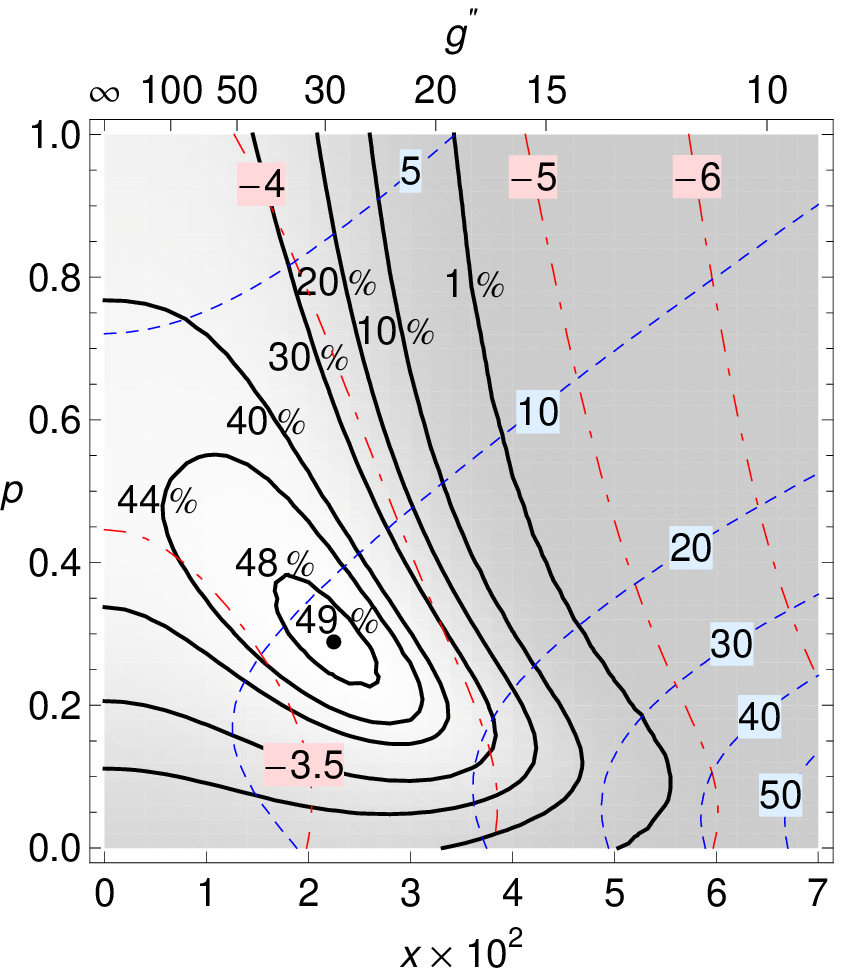}
\caption{\label{fig:ContourMapsScalar125GeV}(color online)
         The same as in Fig.~\ref{fig:ContourMapsHiggsless}
         except that the 125-GeV scalar case
         ($M_H=125\unit{GeV}$) is considered here.
         The left and right panels display the $\Lambda=1\unit{TeV}$
         and $\Lambda=2\unit{TeV}$ cases, respectively.
         }
\end{figure*}

In Figs.~\ref{fig:ContourMapsHiggsless} and
\ref{fig:ContourMapsScalar125GeV}, there are the contour maps of
backings based on the Set-B $\{\Delta L, \Delta R\}$ fits of the
Higgsless top-BESS model and the 125-GeV scalar top-BESS model,
respectively. In this case, the best-fit values of $\{\Delta L,
\Delta R\}$ and their backing depends on the given $x$ and $p$.
Figs.~\ref{fig:ContourMapsHiggsless} and
\ref{fig:ContourMapsScalar125GeV} depict contours of constant
backing in the $(x,p)$ parametric space. At the same time, the
graphs contain the grid lines of constant best-fit values of
$\Delta L$ and $\Delta R$: $\Delta L(x,p)=\mathrm{const}$ and
$\Delta R(x,p)=\mathrm{const}$. The grid can be used to read off
the best-fit values of $\Delta L$ and $\Delta R$ for a given pair
of $x$ and $p$.

In Fig.~\ref{fig:ContourMapsHiggsless}, there are three Higgsless
graphs corresponding to $\Lambda=0.3$, $1$, and $2\unit{TeV}$. We
can see how the low-energy precision data push $g''$ to infinity
and $p$ close to zero. Regarding $g''$, the situation gets worse
as $\Lambda$ grows.

Fig.~\ref{fig:ContourMapsScalar125GeV} shows two 125-GeV scalar
graphs, one for $\Lambda=1\unit{TeV}$, the other for
$\Lambda=2\unit{TeV}$. When compared with the Higgsless case, the
preferred value of $g''$ moved to about $30$ and the data support
has risen. The preferred value of $p$ differs from zero.
Nevertheless, it still clearly suggests that the vector resonance
interaction to the right bottom quark should be weaker than to the
right top quark.

\subsection{Low-energy limits in the \boldmath{$(\Delta L,\Delta R)$} space
in the 125-GeV scalar case} \label{sec:LELimitsInDLDRspace}

For the case of the 125-GeV scalar top-BESS model we have
calculated allowed regions in the $(\Delta L,\Delta R)$ parameter
space based on the two-parameter fits of the Set-B observables
($\mathrm{d.o.f.}=3$). Of course, the fitting parameters are
$\Delta L$ and $\Delta R$, the values of $g''$, $p$, and $\Lambda$
are fixed.

\begin{figure}
\includegraphics[scale=0.59]{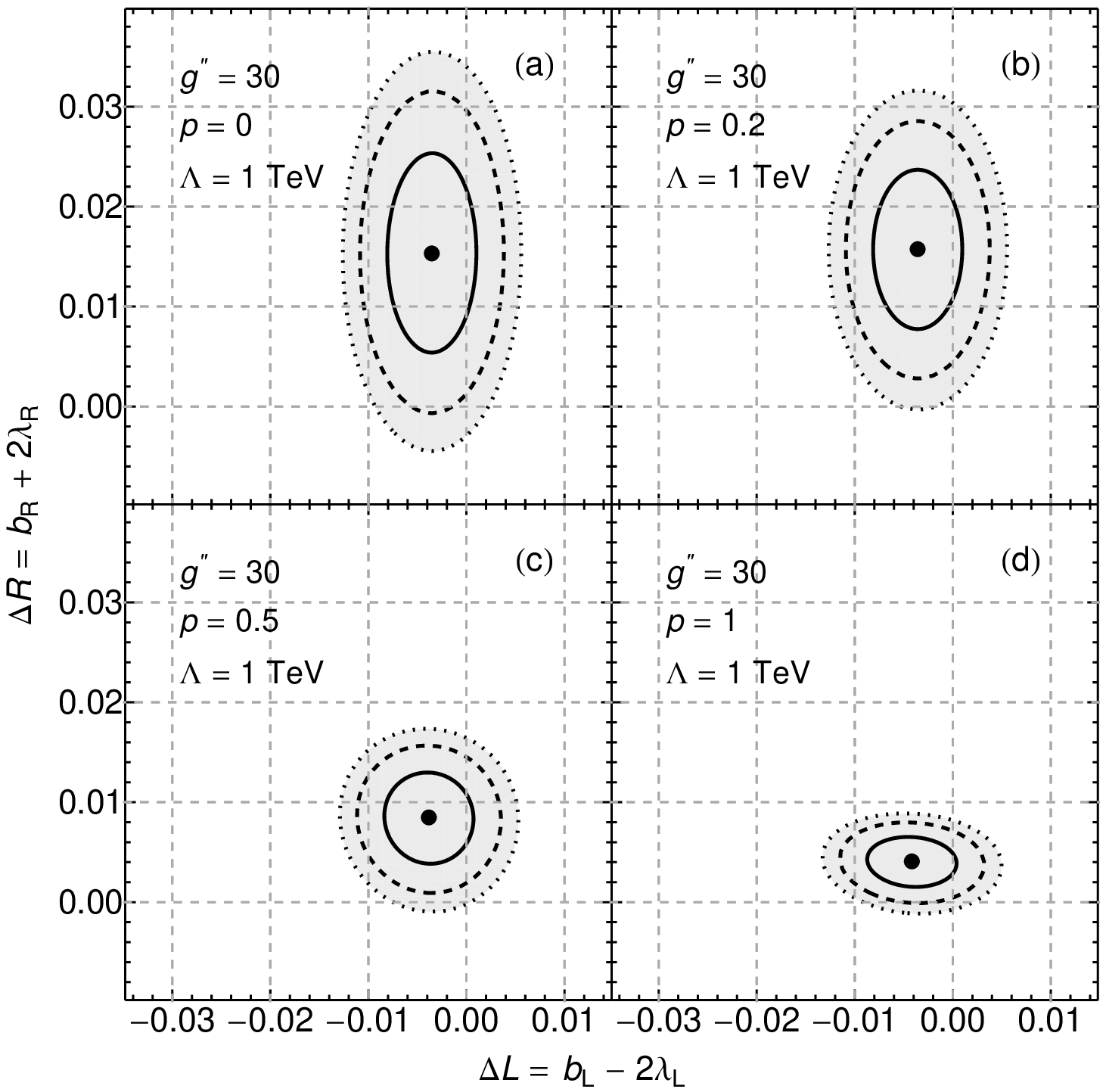}
\caption{\label{fig:CL-gpp:30-p:All-Lamb:1}
 90\% C.L. (solid line), 95\% C.L. (dashed line),
 and 99\% C.L. (dotted line) allowed regions in
 the $(\Delta L, \Delta R)$ parameter space.
 The regions are derived from the two-parameter fit
 of the Set-B observables ($\mathrm{d.o.f.}=3$)
 by the 125-GeV scalar top-BESS models
 with $g''=30$, $\Lambda=1$~TeV, and $p=0$ (a),
 $0.2$ (b), $0.5$ (c), and $1$ (d).
 The fitting parameters are $\Delta L$ and $\Delta R$
 and their best-fit values are indicated by the dots.
 }
\end{figure}

In Fig.~\ref{fig:CL-gpp:30-p:All-Lamb:1} we show the 90\%, 95\%,
and 99\% confidence level regions around the best-fit points with
$g''=30$, $\Lambda=1\unit{TeV}$, and $p=0$, $0.2$, $0.5$, and $1$.
The backings of these parameter points are about $12\%$, $49\%$,
$29\%$, and $14\%$, respectively. Since $\Delta L$ is
predominantly related to a different observable ($\Gamma_b$) than
$\Delta R$ and $p$ ($\epsilon_1$, $\Gamma(\bgs)$), it is not
unexpected that the changes in $p$ affect $\Delta R$ only. In all
four displayed cases the $95\%$~C.L.\ allowed interval for $\Delta
L$ reads $(-0.011,0.004)$. The $95\%$~C.L.\ allowed interval for
$\Delta R$ is $(-0.001,0.032)$ when $p=0$. It shrinks to $\Delta
R\in(0.000,0.008)$ when $p=1$.

\begin{figure}
\includegraphics[scale=0.59]{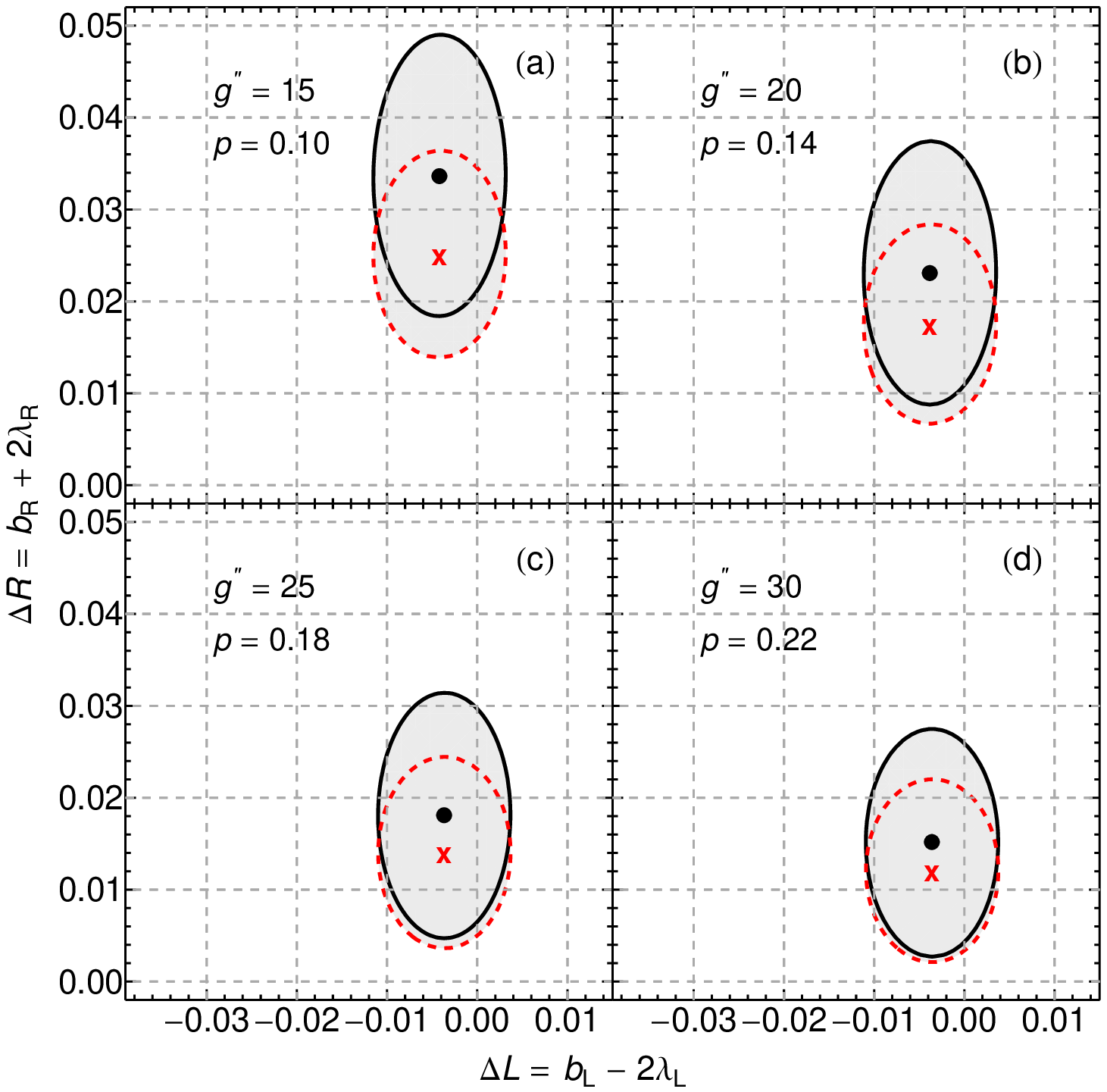}
\caption{\label{fig:CL-gpp:10.20.30-p:all-Lamb:1}(color online)
 95\% C.L.\ allowed regions in
 the $(\Delta L, \Delta R)$ parameter space derived from the same
 fit as in Fig.~\ref{fig:CL-gpp:30-p:All-Lamb:1} except for
 the values of the fixed parameters.
 The black solid contours correspond to $\Lambda=1\unit{TeV}$,
 the red dashed ones to $\Lambda=2\unit{TeV}$.
 The values of the fixed parameters $g''$ and $p$ are, respectively,
 $15$ and $0.10$ (a), $20$ and $0.14$ (b),
 $25$ and $0.18$ (c), $30$ and $0.22$ (d).
 The best-fit values of $\Delta L$ and $\Delta R$ are
 indicated by the black dot and red cross
 for $\Lambda=1\unit{TeV}$ and $\Lambda=2\unit{TeV}$, respectively.
 }
\end{figure}

Fig.~\ref{fig:CL-gpp:10.20.30-p:all-Lamb:1} shows the 95\% C.L.\
allowed regions in the $(\Delta L, \Delta R)$ parameter space when
$p$ assumes the values $0.10$, $0.14$, $0.18$, and $0.22$. These
are the $p$ values with the highest backings for $g''=15$, $20$,
$25$, and $30$, respectively. The corresponding backings are about
$18\%$, $42\%$, $49\%$, and $49\%$. The allowed regions are shown
for the cut-off scales $\Lambda=1\unit{TeV}$ and
$\Lambda=2\unit{TeV}$. As far as the allowed regions are
concerned, $\Delta L$ falls within $(-0.012, 0.004)$,
independently of the other parameter values. For
$\Lambda=1\unit{TeV}$, the $\Delta R$ limits read $(0.018, 0.049)$
when $g''=15$, $(0.009, 0.037)$ when $g''=20$, $(0.005, 0.032)$
when $g''=25$, and $(0.002, 0.028)$ when $g''=30$. For
$\Lambda=2\unit{TeV}$, the $\Delta R$ limits read $(0.014, 0.036)$
when $g''=15$, $(0.007, 0.028)$ when $g''=20$, $(0.004, 0.024)$
when $g''=25$, and $(0.002, 0.022)$ when $g''=30$.

\subsection{Low-energy regions vs the Death Valley effect}
\label{sec:LEvsDV}

In the paper~\cite{tBESS}, we had introduced and analyzed the
so-called Death Valley (DV) effect. The DV is a region in the
$(b_L,b_R)$ parameter space where the interplay of the direct and
indirect couplings of the vector triplet with fermions can
diminish or even zero a particular top/bottom quark channel decay
width of the vector resonance. Thus, it might happen that even
though the direct couplings of the vector resonance to the top
and/or bottom quark --- proportional to $b_{L,R}$ --- are
non-trivial the resonance will not decay through the given quark
channel. Or, the particular decay will be suppressed below the
value that would be implied by the indirect couplings alone.

We have calculated the DV regions for $M_V=1\unit{TeV}$ and for
the parameter values of
Figs.~\ref{fig:CL-gpp:10.20.30-p:all-Lamb:1}a and
\ref{fig:CL-gpp:10.20.30-p:all-Lamb:1}d . That is, $(g'',p)=(15,\,
0.10)$ in the former case, and $(30,\, 0.22)$ in the latter one.
For each case, the DV's in three decay channels of the vector
resonance triplet have been found. The channels are
$V^\pm\rightarrow t\bar{b}/\bar{t}b$, $V^0\rightarrow b\bar{b}$,
and $V^0\rightarrow t\bar{t}$.

\begin{figure*}
\centering
\includegraphics[scale=0.75]{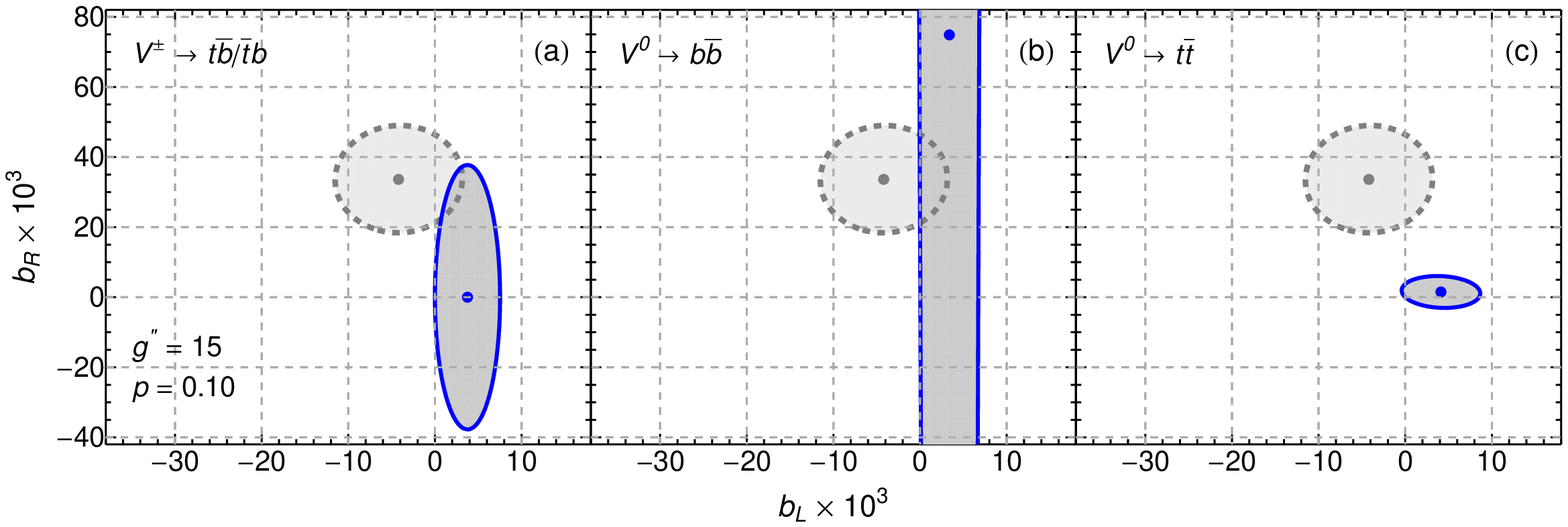}
\caption{\label{fig:DV-gpp:15-p:0.10}(color online)
 The DV regions (dark-shaded areas with the blue solid boundary)
 of the $M_V=1\unit{TeV}$ vector resonance for three decay channels:
 (a) $V^\pm\rightarrow t\bar{b}/\bar{t}b$,
 (b) $V^0\rightarrow b\bar{b}$, and
 (c) $V^0\rightarrow t\bar{t}$.
 The blue dots inside the DV's indicate the point of
 no decay for the particular channels.
 The DV regions are calculated for $g''=15$ and $p=0.10$.
 The corresponding $95\%$~C.L.\ EWPD contours for $\Lambda=1\unit{TeV}$
 and $\lambda_L=\lambda_R=0$ are superimposed to the graphs as
 the regions with the gray dashed boundaries.
 The gray dots inside the EWPD regions indicate the point
 with the highest backing. Note that the DV for
 the $V^0\rightarrow b\bar{b}$ channel exceeds the displayed
 range of the $b_R$ axis. The complete DV region has an oval shape
 centered at the blue dot. Its lower and upper ends
 are found at $b_R=-0.272$ and $b_R=0.422$, respectively.
 }
\end{figure*}

In Fig.~\ref{fig:DV-gpp:15-p:0.10}, the DV regions of the three
decay channels corresponding to
Fig.~\ref{fig:CL-gpp:10.20.30-p:all-Lamb:1}a parameters ($g''=15$
and $p=0.1$) are depicted. In each of the three graphs, there are
the $95\%$~C.L.\ electroweak precision data (EWPD) contours of
Fig.~\ref{fig:CL-gpp:10.20.30-p:all-Lamb:1}a for
$\Lambda=1\unit{TeV}$ superimposed. The low-energy limits apply to
the combination of $b$'s and $\lambda$'s ($\Delta L$ and $\Delta
R$) rather than to the parameters alone. The low-energy limits
depicted in the $b_R-b_L$ graphs correspond to
$\lambda_L=\lambda_R=0$. Nevertheless, by choosing non-zero values
for $\lambda_{L,R}$ the low-energy contours get shifted around the
$(b_L,b_R)$ parameter space. Various values of $\lambda$'s can
result either in no overlap of the low-energy regions with the
DV's or in maximal overlap of the two areas.

\begin{figure*}
\includegraphics[scale=0.75]{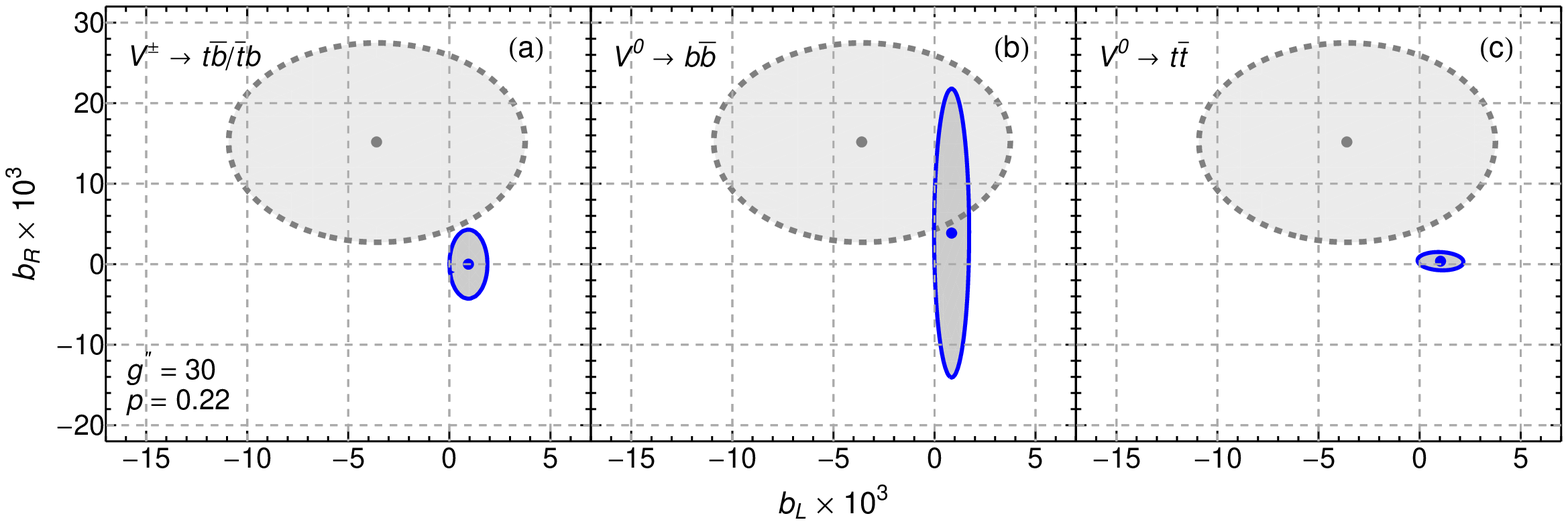}
\caption{\label{fig:DV-gpp:30-p:0.22}(color online)
 The same as in Fig.~\ref{fig:DV-gpp:15-p:0.10} except
 for different values of $g''$ and $p$. In this case,
 $g''=30$ and $p=0.22$.
 }
\end{figure*}

Fig.~\ref{fig:DV-gpp:30-p:0.22} displays the same contents as
Fig.~\ref{fig:DV-gpp:15-p:0.10}, except for different values of
the parameters $g''$ and $p$. Here, the values of the parameters
correspond to Fig.~\ref{fig:CL-gpp:10.20.30-p:all-Lamb:1}d:
$g''=30$ and $p=0.22$. There are also the $95\%$~C.L.\ electroweak
precision data contours of
Fig.~\ref{fig:CL-gpp:10.20.30-p:all-Lamb:1}d for
$\Lambda=1\unit{TeV}$ and $\lambda_L=\lambda_R=0$ superimposed in
Fig.~\ref{fig:DV-gpp:30-p:0.22}.

As we can see in Figs.~\ref{fig:DV-gpp:15-p:0.10} and
\ref{fig:DV-gpp:30-p:0.22} the DV areas are more or less
comparable in size with the EWPD regions. In addition, both
structures are located not far away from each other when
$\lambda_L=\lambda_R=0$. Hence, there are reasonable values of the
$\lambda$ parameters for which the significant part of an EPWD
region falls inside the DV.

There might be new physics materialized through the existence of
the new vector resonances as well as non-zero values of the $b$
parameters, yet it does not have to reveal itself in an
experiment. If the actual values of the $b$ parameters fell in the
DV it would make the detection and study of the new vector
resonance more difficult. In particular, thanks to the indirect
mixing-induced coupling the vector resonance can be produced and
studied in the Drell-Yan processes at the LHC or in the s-channel
at a future electron-positron collider~\cite{newBESS,DY}. As we
demonstrated in \cite{tBESS} the signal of the vector resonance in
the top and bottom decay channels can be diminished or hidden by
the negative interference between the direct and indirect
couplings.

\section{Conclusions}
\label{sec:Conclusions}

The top-BESS effective Lagrangian describing the phenomenology of
the $SU(2)$ vector resonance triplet was extended to study the
impact of a newly discovered 125-GeV boson on the low-energy
limits and the statistical support for the model. We opted for the
assumption that the boson was a scalar with the SM-like
parameters. The $\chi^2$ analysis based on the epsilon
pseudo-observables, $\Gamma_b(Z\rightarrow b\bar{b}+X)$, and
BR$(\bgs)$, resulted in finding the best-fit values and the
backings for the vector resonance gauge coupling $g''$ and for
three parameters responsible for the direct vector resonance
couplings to the top and bottom quarks, $\Delta L=b_L-2\lambda_L$,
$\Delta R=b_R+2\lambda_R$, and $p$. The analysis was performed
assuming TeV values of the vector resonance mass.

Without the scalar resonance the low-energy data pushed the
preferred value of $g''$ to infinity which would effectively
amount to removing the vector resonance from the model's spectrum.
In addition, when $g''\gtrsim 30$ we enter the non-perturbative
regime in which our conclusions are questionable. The optimal
value of $p$ was approaching zero which would amount to turning
off the vector resonance coupling to the right bottom quark. The
maximum data backing for this scenario did not exceed $30\%$.

After the scalar resonance had been added the situation improved
in a couple of ways. The most preferred value of $g''$ decreased
to about 30 and its backing grew to about $50\%$. The best-fit
value of $p$ ranges between $0.2$ and $0.3$ for $M_V=1\unit{TeV}$
and $2\unit{TeV}$, respectively. This result suggests the
preferred role of the top quark in physics responsible for ESB.
Recall that in the top-BESS model the $V^\pm t_Rb_R$ vertex is
suppressed by the factor $p$ while the suppression factor of the
$V^0b_Rb_R$ vertex relative to $V^0t_Rt_R$ is $p^2$.

We calculated the $95\%$ confidence level allowed regions around
the best-fit values of $\Delta L$ and $\Delta R$ when $g''$ and
$p$ had been fixed. This imitates the situation when the concerned
parameters were constrained by theoretical assumptions. We chose
$g''=15$, $20$, $25$, and $30$ and $p=0.10$, $0.14$, $0.18$, and
$0.22$, respectively. The values of $p$ were fixed at their
best-fit values for the given $g''$'s. The backings for the
corresponding best-fit values of $(\Delta L,\Delta R)$ were
$18\%$, $42\%$, $49\%$, and $49\%$, respectively, with
$\mathrm{d.o.f.}=3$. For all these settings, the $95\%$~C.L.
interval of $\Delta L$ is $(-0.012,0.004)$. When $M_V=1\unit{TeV}$
the $95\%$~C.L. interval of $\Delta R$ shrinks from
$(0.018,0.049)$ at $g''=15$ to $(0.002,0.028)$ at $g''=30$.
Raising $M_V$ to $2\unit{TeV}$ did not affect the $95\%$~C.L.
interval for $\Delta L$. The $95\%$~C.L. interval for $\Delta R$
changes from $(0.014,0.036)$ at $g''=15$ to $(0.002,0.022)$ at
$g''=30$.

There might be new physics materialized through the existence of
the new vector resonances as well as non-zero values of the $b$
parameters, yet it does not have to reveal itself in an
experiment. Even though there are no direct interactions of the
vector resonance triplet to the light fermions the resonance does
couple to the light fermions thanks to the mixing with the
electroweak gauge bosons. This enables processes with a direct
production of the vector resonance at the LHC and future
electron-positron colliders. However, in the top and bottom decay
channels the signal of the vector resonance can be diminished or
hidden by the negative interference between the direct and
indirect couplings. We calculated the regions of the negative
interference for the studied values of the top-BESS parameters and
found that they are often comparable in size and close in position
with the $95\%$~C.L. regions. As a general tendency, the relative
size of the DV's with respect to the low-energy allowed areas
shrinks as the value of $g''$ grows.

\begin{acknowledgments}
We would like to thank Ivan Melo for useful discussions. The work
of M.G.\ and J.J.\ was supported by the Research Program
MSM6840770029 and by the project International Cooperation
ATLAS-CERN of the Ministry of Education, Youth and Sports of the
Czech Republic. M.G.\ was supported by the Slovak CERN Fund. J.J.\
was supported by the NSP grant of the Slovak Republic. We would
also like to thank the Slovak Institute for Basic
Research for their support.\\
\end{acknowledgments}

\appendix

\section{Experimental values}
\label{app:ExpValues}

In our analyses we have used the experimental values of the
epsilon pseudo-observables obtained from a fit to all LEP-I and
SLD measurements including the combined preliminary measurement of
the $W$-boson mass \cite{EpsilonData}
\begin{eqnarray}
  \epsilon_1^{\mathrm{exp}} &=&(+5.4\phantom{4} \pm 1.0) \times 10^{-3},
  \label{epsExp1}\\
  \epsilon_2^{\mathrm{exp}} &=&(-8.9\phantom{4} \pm 1.2) \times 10^{-3},
  \label{epsExp2}\\
  \epsilon_3^{\mathrm{exp}} &=&        (+5.34 \pm 0.94) \times 10^{-3},
  \label{epsExp3}\\
  \epsilon_b^{\mathrm{exp}} &=&(-5.0\phantom{4} \pm 1.6) \times 10^{-3},
  \label{epsExpb}
\end{eqnarray}
with the correlation matrix
\begin{eqnarray}
\rho^\epsilon =
\left(%
\begin{array}{cccc}
  1.00 &  \phantom{-}0.60 & \phantom{-}0.86 &  \phantom{-}0.00 \\
  0.60 &  \phantom{-}1.00 & \phantom{-}0.40 & -0.01 \\
  0.86 &  \phantom{-}0.40 & \phantom{-}1.00 &  \phantom{-}0.02 \\
  0.00 & -0.01 & \phantom{-}0.02 &  \phantom{-}1.00 \\
\end{array}%
\right).
  \label{epsExpCM}
\end{eqnarray}

The value of the $Z\rightarrow b\bar{b}$ decay width
\begin{equation}
 \Gamma_b^{\mathrm{exp}} = (0.3773 \pm 0.0013)\ \mathrm{GeV}
\end{equation}
has been obtained from the experimental values~\cite{PDG2010}
\begin{eqnarray}
 \mbox{BR}(Z\rightarrow b\bar{b})^{\mathrm{exp}} &=& (0.1512\pm 0.0005),
 \\
 \Gamma_{tot}(Z)^{\mathrm{exp}}\;\;\;\;\; &=& (2.4952 \pm 0.0023)~\mbox{GeV}.
\end{eqnarray}
When fitting the set
$\{\epsilon_1,\epsilon_2,\epsilon_3,\Gamma_b\}^{\mathrm{exp}}$ the
correlations between $\Gamma_b$ and $\epsilon_{1,2,3}$ have been
neglected.

For the branching fraction of $\bgs$ we have used the world
average \cite{btosgData} (CLEO, Belle, BaBar)
\begin{equation}
 \mbox{BR}(\bgs)^{\mathrm{exp}} = (3.55 \pm 0.26)\times 10^{-4}.
\end{equation}
We have considered no correlations between $\mbox{BR}(\bgs)$ and
any of the observables
$\epsilon_1,\epsilon_2,\epsilon_3,\epsilon_b,\Gamma_b$.

Below we will complete the list of numerical values that have been
used in the calculations of this paper
\begin{eqnarray}
 \alpha(0)        &=& 1/137.036,
 \\
 \alpha(M_Z^2)    &=& 1/128.91,
 \\
 \alpha_s(M_Z^2)  &=& 0.1184,
 \\
 G_F  &=& 1.166364\times 10^{-5} \ \mathrm{GeV^{-2}},
 \\
 m_b  &=& 4.67 \ \mathrm{GeV},
 \\
 m_t  &=& 172.7 \ \mathrm{GeV},
 \\
 M_Z  &=& 91.1876 \ \mathrm{GeV}.
\end{eqnarray}
Then, using the Eq.~(\ref{eq:s0fromGF}) the numerical value of
$s_0^2$ is
\begin{equation}
 s_0^2 = 0.2311.
\end{equation}

\section{Low-energy top-BESS Lagrangian}
\label{app:LEtBESS}

To obtain the low-energy limits on the top-BESS parameters we have
derived the low-energy Lagrangian (LE-tBESS) by integrating out
the vector triplet from the top-BESS Lagrangian with the 125 GeV
scalar resonance (\ref{eq:ExtendLag}). It proceeds by taking the
limit $M_{triplet}\rightarrow\infty$, while $g''$ is finite and
fixed, and by substituting the equation of motion (EofM) for the
triplet fields obtained under these conditions. Most of the time
the obtained relations shown in this Appendix are identical with
those of the top-BESS Lagrangian without the scalar resonance. We
will point out any differences that will be encountered.

The EofM in the unitarity gauge reads
\begin{equation}
 i\frac{g''}{2}V_\mu^a = \frac{1}{2}(igW_\mu^a+ig'B_\mu\delta^{a3}),
\end{equation}
where $a=1,2,3$.

The gauge boson kinetic terms of the LE-tBESS Lagrangian in the
unitary gauge read
\begin{eqnarray}
  \cL_{kin}^{\mathrm{LE}}(A',Z') &=&
  -\frac{1}{4}(1+z_z)F_{\mu\nu}(Z')F^{\mu\nu}(Z')
  \nonumber\\
  && -\frac{1}{4}(1+z_\gamma)F_{\mu\nu}(A')F^{\mu\nu}(A')
  \nonumber\\
  && +\frac{1}{2}z_{z\gamma}F_{\mu\nu}(Z')F^{\mu\nu}(A'),
  \\
  \cL_{kin}^{\mathrm{LE}}(W^{\pm\prime}) &=&
  -\frac{1}{4}(1+x^2)\left[F_{\mu\nu}^\dagger(W^{+\prime}) F^{\mu\nu}(W^{+\prime})\right.
  \nonumber\\
  && +\left. F_{\mu\nu}^\dagger(W^{-\prime}) F^{\mu\nu}(W^{-\prime})\right],
\end{eqnarray}
where $F_{\mu\nu}(X)\equiv\pard_\mu X_\nu - \pard_\nu X_\mu$, and
\begin{equation}\label{eq:z}
  z_z = \left(\frac{c_{2\theta}}{c_\theta}x\right)^2,\;\;
  z_\gamma = 4s_\theta^2x^2,\;\;
  z_{z\gamma} = -\sqrt{z_zz_\gamma},
\end{equation}
where
\begin{equation}\label{eq:x}
  x\;\equiv\;\frac{g}{g''}.
\end{equation}
The primed fields are the gauge fields of the top-BESS Lagrangian
in the mass eigenstate basis and $s_\theta$, $c_\theta$ are the
elements of the transformations matrix taking $(W_\mu^3,B_\mu)$ to
$(Z_\mu',A_\mu')$.

To obtain the canonical kinetic terms for the LE-tBESS Lagrangian
the fields must undergo the following set of transformation
\begin{eqnarray}
  Z'_\mu &=& \sqrt{\frac{1+z_\gamma}{1+z_\gamma+z_z}} Z_\mu,
  \label{LERenormZ}\\
  A'_\mu &=& \frac{z_{z\gamma}}{\sqrt{(1+z_\gamma)(1+z_\gamma+z_z)}} Z_\mu
            +\frac{1}{\sqrt{1+z_\gamma}} A_\mu,\;\;\;\;\;
  \label{LERenormA} \\
  W^{\pm\prime}_\mu &=& \frac{1}{\sqrt{1+x^2}} W^\pm_\mu.
  \label{LERenormW}
\end{eqnarray}
Then, the masses of the ``low-energy'' electroweak gauge fields
$Z_\mu$ and $W_\mu^\pm$ are given by the expressions
\begin{eqnarray}
 M_{Z}^2 &=& \frac{1+z_\gamma}{1+z_\gamma+z_z}\frac{G^2v^2}{4},
 \label{LEmassZ}\\
 M_{W}^2 &=& \frac{1}{1+x^2}\frac{g^2v^2}{4},
 \label{LEmassW}
\end{eqnarray}
where $G=\sqrt{g^2+g'^2}$. However, it is convenient to choose
$x$, $s_\theta$, and $M_Z$ as input parameters of the LE-tBESS
Lagrangian. Then, the mass of the $W$ boson is a function of these
parameters
\begin{equation}
 M_W = M_Z c_\theta\sqrt{\frac{1+(x/c_\theta)^2}{(1+4s_\theta^2 x^2)(1+x^2)}}.
\end{equation}

Let us turn our attention to the fermion sector of the LE-tBESS
Lagrangian. After integrating out the vector triplet we end up
with the low-energy Lagrangian $\cL_{\mathrm{ferm}}^{\mathrm{LE}}$
in the unitarity gauge expressed in terms of the top-BESS gauge
fields $A_\mu'$, $Z_\mu'$, $W_\mu^{\pm'}$. Eventually, to find the
form of the low-energy fermion interactions the renormalized
fields (\ref{LERenormZ}) -- (\ref{LERenormW}) have to be
substituted in. Then\footnote{To obtain
$\cL_{\mathrm{ferm}}^{\mathrm{LE}}$ for the case of the top-BESS
model without the scalar resonance set the scalar field to zero,
$h=0$, in the Eq.~(\ref{eq:LEfermionLagr}).},
\begin{eqnarray}\label{eq:LEfermionLagr}
 \cL_{\mathrm{ferm}}^{\mathrm{LE}} &=& i\bar{\psi}\dslash \psi -e\bar{\psi}\Aslash Q\psi
 -\frac{G_N}{2}\bar{\psi}\Zslash (C_LP_L+C_RP_R)\psi
 \nonumber\\
 && - \frac{G_C c_\theta}{\sqrt{2}}\bar{\psi}(\Wslash^+\tau^++\Wslash^-\tau^-)
      (D_LP_L+D_RP_R)\psi
 \nonumber\\
 && -(\bar{\psi}_L M_f\psi_R + \mbox{H.c.})\left(1+h/v\right),
\end{eqnarray}
where $\psi$ are fermion $SU(2)$ doublets, $\tau^\pm=\tau^1\pm
i\tau^2$, $P_{L,R}=(1\mp\gamma_5)/2$, $M_f=\diag(m_u,m_d)$ is a
fermion mass matrix\footnote{ Therein, $m_u$ and $m_d$ stand for
the masses of any of the upper and lower components of the fermion
$SU(2)$ doublets, respectively.}, and
\begin{equation}
 G_N = \frac{e}{s_\theta c_\theta}\frac{1+z_\gamma}{\sqrt{1+z_\gamma+z_z}},
 \gap
 G_C = \frac{e}{s_\theta c_\theta}\sqrt{\frac{1+z_\gamma}{1+x^2}}.
\end{equation}
For the light fermions (all SM fermions except the top and bottom
quarks)
\begin{eqnarray}
 C_L &=& 2(T_L^3-s_\theta^2 Q)
                -2s_\theta c_\theta\frac{\sqrt{z_\gamma z_z}}{1+z_\gamma}Q_f,
 \\
 C_R &=& -2s_\theta^2 Q
                -2s_\theta c_\theta\frac{\sqrt{z_\gamma z_z}}{1+z_\gamma}Q_f,
\end{eqnarray}
and $D_L = 1$, $D_R = 0$. In the case of the top and bottom quarks
\begin{eqnarray}
 C_L &=& 2[(1-\Delta L/2) T_L^3-s_\theta^2 Q]
                -2s_\theta c_\theta\frac{\sqrt{z_\gamma z_z}}{1+z_\gamma}Q_f,\;\;\;\;\;\;\;
 \label{eq:CLtopbottomquarks}\\
 C_R &=& 2(P_f\Delta R T_R^3/2-s_\theta^2 Q)
                -2s_\theta c_\theta\frac{\sqrt{z_\gamma z_z}}{1+z_\gamma}Q_f,\;\;\;
 \label{eq:CRtopbottomquarks}
\end{eqnarray}
where $P_t = 1$, $P_b = p^2$, and
\begin{equation}\label{eq:DLRtopbottomquarks}
 D_L = 1-\Delta L/2,\gap D_R = p\; \Delta R/2.
\end{equation}

The following list contains a convenient choice of independent
input parameters for the LE-tBESS Lagrangian
\begin{equation}\label{eq:LEtBESSparams}
  e,s_\theta,x,M_Z,\Delta L,\Delta R,p,\{m_f\},
\end{equation}
where $\{m_f\}$ are the non-negligible fermion masses. The list of
independent input parameters of the top-BESS model, natural from
the point of view of the model construction, reads
\begin{equation}\label{eq:tBESSparams}
  g,g',g'',\alpha,v,b_L,b_R,\lambda_L,\lambda_R,p,\{m_f\}.
\end{equation}
Of course, the parameters (\ref{eq:LEtBESSparams}) are related to
the parameters (\ref{eq:tBESSparams}) of the underlying theory.
Thus, the electric charge relation to the top-BESS parameters
reads
\begin{equation}
 e = \frac{g g' g''}{\sqrt{(g g'')^2+(g'g'')^2+(2g g')^2}}.
\end{equation}
Further,
\begin{equation}
 s_\theta = \frac{g'}{\sqrt{g^2+g^{\prime 2}}},
 \gap
 x = \frac{g}{g''}.
\end{equation}
The $Z$-boson mass relation is given by
\begin{equation}
 M_Z^2 = \frac{(g g'')^2+(g'g'')^2+(2g g')^2}{g^2+g^{\prime 2}+g^{\prime\prime 2}}
         \;\frac{v^2}{4}.
\end{equation}
Finally,
\begin{equation}\label{eq:DeltaLR}
 \Delta L = b_L-2\lambda_L,
 \gap
 \Delta R = b_R+2\lambda_R.
\end{equation}

In the LE-tBESS model the Fermi coupling $G_F$ is related to $e$,
$s_\theta$, $M_Z$, and $x$
\begin{equation}\label{eq:GFinLEtBESS}
   \frac{G_F}{\sqrt{2}}=\frac{1}{2}\left(\frac{e}{2s_\theta c_\theta M_Z}\right)^2
                        \frac{(1+4s_\theta^2 x^2)^2}{1+(\frac{x}{c_\theta})^2}.
\end{equation}
If the values of $G_F$, $e(M_Z)$, and $M_Z$ are fixed by
measurements then we can obtain $s_\theta$ as a function of $x$:
$s_\theta(x;e(M_Z),M_Z,G_F)$. However, as mentioned in Section
\ref{sec:pseudoobservables}, $G_F$ can be replaced by $s_0$ using
the SM relation (\ref{eq:s0fromGF}). Comparing
(\ref{eq:GFinLEtBESS}) with (\ref{eq:s0fromGF}), the following
relation between $s_{\theta}$ and $s_0$ must hold
\begin{equation}\label{eq:s0c0vssThetacTheta}
 s_0 c_0 = s_\theta c_\theta \frac{\sqrt{1+(\frac{x}{c_\theta})^2}}{1+4s_\theta^2 x^2}.
\end{equation}
Note that using the $s_0$ parameter for determination of
$s_\theta$ eliminates, beside $G_F$, also $e(M_Z)$ and $M_Z$; or,
in other words, for the given value of $s_0$, $s_\theta$ is a
function of $x$ only, $s_\theta(x;s_0)$. The acceptable solution
of the implicit equation for $s_\theta$
(\ref{eq:s0c0vssThetacTheta}) reads
\begin{equation}\label{eq:sTheta}
 s_\theta = \sqrt{\frac{1+c_{40}x^2-(1+x^2)\sqrt{1-s_{20}^2\left(\frac{1+2x^2}{1+x^2}\right)^2}}
                   {2(1+4s_{20}^2x^4)}}
\end{equation}
assuming $c_\theta >0$ and defining $s_{20}\equiv 2s_0c_0$,
$c_{40}\equiv c_{20}^2-s_{20}^2$. Further, as a consequence of
(\ref{eq:s0c0vssThetacTheta}) the couplings $G_N$ and $G_C$ can be
expressed in the way that will prove useful for deriving LE-tBESS
contributions to the anomalous fermion couplings
\begin{equation}
 G_N = \frac{e}{s_0c_0},\gap
 G_C c_\theta = \frac{e}{s_0}
 \left(\frac{s_0}{s_\theta}\sqrt{\frac{1+z_\gamma}{1+x^2}}\right).
\end{equation}

Given the measured values of $G_F$, $e(M_Z)$, and $M_Z$ (or,
alternatively, $s_0$, $\alpha(M_Z)$, and $M_Z$), the top-BESS
parameters $g$ and $g''$ depend on $x$ in the following way
\begin{eqnarray}
  g(x) &=& \frac{e(M_Z)}{\sqrt{2}s_0c_0}\left[1+x^2+\right.
  \nonumber\\
       &&  \left.\sqrt{c_{20}^2 (1+x^2)^2 - s_{20}^2x^2(2+3x^2)}\right]^{1/2},
  \label{eq:gx}\\
  g''(x) &=& \frac{g(x)}{x}.
  \label{eq:gppx}
\end{eqnarray}
The leading terms of the series expansion of $g$ in $x$ around
$x=0$ reads
\begin{equation}
  g(x) = \frac{e}{s_0}\left[1+\left(1-\frac{1}{2c_{20}}\right)x^2
         +\cO(x^4)\right].
\end{equation}

We often wish to find the value of $x$ which corresponds to a
given $g''$. In other words, we need a formula inverse to
(\ref{eq:gppx}). It reads
\begin{equation}
 x(\eta) = \frac{\eta}{\sqrt{2}}
          \left(1-\eta^2-\sqrt{1-\eta^2-s_{20}^2}\right)^{-1/2},
\end{equation}
where
\begin{equation}
 \eta = \frac{\sqrt{16\,\pi\,\alpha(M_Z)}}{g''} = \frac{2\,e(M_Z)}{g''}
        \approx \frac{0.624}{g''}.
\end{equation}
The leading terms of the series expansion of $x(\eta)$ in $\eta$
around $\eta=0$ read
\begin{equation}
 x = \frac{\eta}{2\,s_0}
     \left[1+\frac{2\,c_{20}-1}{2\, c_{20}}\left(\frac{\eta}{2\,s_0}\right)^2
     + \cO(\eta^4)\right],
\end{equation}
where $\eta/(2\,s_0) = e(M_Z)/(s_0 \,g'')$.

\section{The \boldmath{$\chi^2$}-test}
\label{app:chi2test}

In this Appendix we summarize the relations used in the paper for
statistical calculations concerning the $\chi^2$-test.

First of all, the $\chi^2$-function is defined as
\begin{equation}
 \chi^2 = \sum_{i=1}^n\sum_{j=1}^n
        (O_i-O_i^{exp})[(\sigma^2)^{-1}]_{ij}(O_j-O_j^{exp}),
 \label{chi2}
\end{equation}
where $O_j^{exp}$ is the measured value of an observable, $O_j$ is
its value predicted by theory, and $\sigma^2$ is the covariance
matrix
\begin{equation}
 (\sigma^2)_{ij}=\sigma_i\rho_{ij}\sigma_j,
\end{equation}
$\sigma_i$ is the standard deviation of an observable $O_i$, and
$\rho_{ij}$ is the correlation matrix.

Usually, the theoretical prediction depends on several free
parameters $p_i$, $O_j(p_1,\ldots,p_k)$. The best fit of the given
theory to the given set of measured observables
$\{O_1^{exp},\ldots,O_n^{exp}\}$ is provided by such values of
$\{p_1,\ldots,p_k \}$ that minimize the function $\chi^2$.

The statistical support for the best-fit values of the fitting
parameters is given by the \textit{backing}
\begin{equation}
 \mathrm{Backing} = \int_{\chi^2_{min}}^\infty f(z;\mathrm{d.o.f.})\; dz,
 \label{backing}
\end{equation}
where $f(z;\mathrm{d.o.f.})$ is the probability density
distribution of $\chi^2$ for $\mathrm{d.o.f.} = n-k$, and
$\chi^2_{min}$ is the global minimum of $\chi^2$.

The probability --- the confidence level --- that the true values
of the free parameters lie within the region of the
$(p_1,\ldots,p_k)$ parameter space for which
\begin{equation}
  \chi^2(p_1,\ldots,p_k) - \chi^2_{min} \leq \Delta\chi^2
 \label{deltachi2}
\end{equation}
is equal to
\begin{equation}
 \mathrm{C.L.} =   \int_0^{\Delta\chi^2} f(z;k) \;dz =
               1-\int_{\Delta\chi^2}^\infty f(z;k)\;dz.
 \label{intCL}
\end{equation}
For the reader's convenience we provide Table~\ref{tab:CL}
relating some values of $\mathrm{C.L.}$ and $\Delta\chi^2$ for
$k=1$, $2$, $3$, and~$4$.

\begin{table}
\caption{The values of $\Delta\chi^2$ for selected
    $\mathrm{C.L.}$ when $k=1$, $2$, $3$, and $4$.
    }
 \label{tab:CL}
\begin{ruledtabular}
\begin{tabular}{ccccc}
  $k$ & 1 & 2 & 3 & 4 \\
  \hline
  C.L. (\%) & \multicolumn{4}{c}{$\Delta\chi^2$} \\
  \hline
  90 & 2.71 & 4.61 & \phantom{1}6.25 & \phantom{1}7.78  \\
  95 & 3.84 & 5.99 & \phantom{1}7.81 & \phantom{1}9.49  \\
  99 & 6.63 & 9.21 & 11.34           & 13.28
\end{tabular}
\end{ruledtabular}
\end{table}



\end{document}